\documentclass[
        parskip = half, %
        twoside,
        footinclude = false, %
        ]{scrartcl}

\usepackage{todonotes}

\usepackage{ifthen} %

\newboolean{boldeqnitalic}
\setboolean{boldeqnitalic}{true}
\usepackage{stylefiles/defdb1} %

\usepackage{stylefiles/rublkm_paper} %

\usepackage{tikz}
\usetikzlibrary{shapes,arrows,arrows.meta,positioning}

\usepackage{siunitx}

\usepackage{xcolor,pifont}
\newcommand*\colourcheck[1]{%
  \expandafter\newcommand\csname #1check\endcsname{\textcolor{#1}{\ding{52}}}%
}
\colourcheck{green}

\newcommand*\colourquestionmark[1]{%
  \expandafter\newcommand\csname #1questionmark\endcsname{\textcolor{#1}{\textbf{?}}}%
}
\colourquestionmark{blue}

\newcommand*\colourcross[1]{%
  \expandafter\newcommand\csname #1cross\endcsname{\textcolor{#1}{\ding{56}}}%
}
\colourcross{red}

\makeatletter
\newcommand*{\authormark}{}
\newcommand*{\markauthor}[1]{%
  \renewcommand{\authormark}{#1}%
  \ignorespaces
}
\newcommand*{\titlemark}{}
\newcommand*{\marktitle}[1]{%
  \renewcommand{\titlemark}{#1}%
  \ignorespaces
}
\makeatother

\markauthor{Dennis Ogiermann et al.}
\marktitle{Voltage-Modulated Piezo1 Ion Channel Model}

\clearpairofpagestyles
\rohead{\pagemark}
\rehead{\pagemark}

\lohead{\titlemark}
\lehead{\authormark}

\KOMAoptions{
   headsepline = true
}
\setkomafont{pagehead}{\normalfont\normalcolor\small}

\newcommand{\figpath}{figures}       %

\begin{document}

\include{sec_00_title}

\clearpage
\setcounter{page}{1}
\begin{center}

{\LARGE A Simple Voltage-Modulated Markov Chain Model for the Piezo1 Ion Channel to Investigate Electromechanical Pacing}

\vspace{5mm}

Dennis Ogiermann$^{1}$, Abdulaziz Mohamed$^{1}$, Luigi E. Perotti$^{2\star}$, Daniel Balzani$^{1}$

\vspace{3mm}

{\small $^1$Chair of Continuum Mechanics, Ruhr-Universität Bochum,\\ 
Universitätsstraße~150, 44801~Bochum, Germany}\\[3mm]

{\small $^2$Department of Mechanical and Aerospace Engineering, University of Central Florida\\
Orlando, FL 32816, USA}\\[3mm]

\vspace{3mm}

{\small ${}^{\star}$E-mail address of corresponding author: Luigi.Perotti@ucf.edu}

\vspace{10mm}

\begin{minipage}{15.0cm}

\textbf{Abstract}\hspace{3mm}
Piezo1 ion channels are voltage-modulated, stretch-activated ion channels involved in a variety of important physiological and pathophysiological processes, as for example cardiovascular development and homeostasis.
Since its discovery, it has been known that this type of ion channel desensitizes when exposed to stretch.
However, recent experiments on Piezo1 ion channels have uncovered that their stretch response is qualitatively different when exposed to positive electrochemical driving forces, where the desensitization is reset.
In this work, we propose a novel voltage-modulated mathematical model of Piezo1 based on a continuous-time Markov chain. 
We show that our Piezo1 model is able to quantitatively reproduce a wide range of experimental observations.
Furthermore, we integrate our new ion channel model into the Mahajan-Shiferaw ventricular cardiomyocyte model to study the effect of electromechanical pacing at the cellular scale.
This integrated cell model is able to qualitatively reproduce some aspects of the experimental observations regarding the rate-dependence of electromechanical pacing protocols. 
Our studies suggest that the Piezo1 ion channel is an important component that significantly contributes to the electromechanical coupled response of cardiomyocytes.
\end{minipage}
\end{center}

\medskip{}

\medskip{}
{
\textbf{Key Points}
\begin{itemize}
    \item PIEZO ion channels are voltage-modulated, mechanically gated ion channels involved in a large variety of mechanically regulated physiological processes and diseases.
    \item Recent experiments on Langendorff perfused rabbit hearts by A. Quinn and P. Kohl [2016] suggest a non-trivial relation between the number of captured mechanical stimuli and the electromechanical pacing protocol.
    \item We present a novel thermodynamically consistent in silico model of the Piezo1 ion channel with electromechanical gating that can reproduce a large variety of experimental observations during combined exposure to electrical and mechanical stimuli.
    \item The new ion channel model is integrated into the well-established Mahajan-Shiferaw rabbit ventricular cardiomyocyte model to study the interaction during normal heart beat and during electromechanical pacing protocols.
    \item Our in silico studies suggest that the Piezo1 ion channel alone may not be sufficient to explain the experimental observations made by A. Quinn and P. Kohl.
\end{itemize}
}

\section{Introduction}

The computational analysis of physiological and pathophysiological heart function has been of increasing interest due to the large potential for improving medical intervention as well as enhancing the fundamental understanding of the underlying processes~\cite{PerKriBorEnnKlu:2015:rsv,CarBueMinZemRod:2016:Hva,PonPerLiuQuWeiEnnKluGar:2016:ehf,DenArePraCalTra:2016:fsa,PasBriLuRohHerGalGreBueRod:2017:hsd,SanDAMafCaiPriKraAurPot:2018:sav,GraDobHei:2019:cmw,AroAliTra:2019:rpa,OgiBalPer:2021:ema,CamLawDroMinWanGraBurRod:2020:Iva,MosWulLewHorPerStrMenKraOdeSee:2022:cmr,OgiBalPer:2023:egh}. 
Especially the electromechanical coupling has attracted particular interest since it dominates the active contraction behavior~\cite{GokMenKuh:2014:ghm,GerDedQuaGerDedQua:2019:mas,GerSchFroLinKovMosWulSeeWieLoe:2021:EWD,AugGseKarWilPriLumVigPla:2021:cep,FedPieRegSalAfrBucZinDedQua:2023:cbd}. 
Interestingly, the mechano-electric feedback has been given less attention, although it is important for many processes in the heart which can be linked to clinically relevant questions. 
For instance, this back-coupling of mechanics onto the chemical processes governs adaptive processes, including pathophysiological growth and homeostasis~\cite{MaivanMol:2013:mbp,Bac:2022:plm}{, and plays a role in arrhythmic processes and commotio cordis~\cite{PeyNerKoh:2016:cmi}}. 
Another example is Cardiopulmonary resuscitation (CPR), a standard emergency procedure~\cite{OlaManPerAviBroCasChuConCouEscHatHunKudLimNisRisSemSmiSmyVaiNolHazMorSvaRafKuzGraDeeSmiRaj:2020:abl,PerGraSemOlaSoaLotVanMadZidMenBosGreMonSvaNolAinAkiAlfAndAttBarBauBehBeiBiaBinBloBocBorBosBotBreBriBurCarCarCarCasCasChrCimClaConCouCroDeDeDeDeaDelDirDjaDjaDruEldErsFriGenGeoGoeGonGraGraGreHanHasHayHelHenHerHinHofHunJohKhaKlaKopKreKuzLauLilLipLocLotLulMaaMacMadMarMasMenMeyMonMorMouMpoNikNolOlaOliPaaPelPerPflPitPooRafRenRisRoeRosRudSafSanSanSarScaSchSchSchSemShaSinSkaSkrSmySoaSvaSzcTacTagTeThiTjeTreTruTruTurUrlVaaVanVanWilWneWylYeuZid:2021:erc} in which fast and deep chest compressions are applied in subjects suffering cardiac arrest to maintain a minimal level of blood circulation until more advanced care can be administered. 
Here, optimized procedure protocols may be identified by additionally exploiting the active contraction response of the heart that may be triggered mechanically. 
Furthermore, existing medical procedures also exploit this back-coupling. 
For instance, approaches such as precordial thump~\cite{PenTayLow:1970:ctr} and percussion pacing~\cite{Sch:1920:vss} aim at conditioning heart pacing using mechanical stimuli. 
The European Resuscitation Society (ERS) recommends that percussion pacing can be attempted during bradycardia, if no pacing equipment is available and atropine treatment is ineffective~\cite{PerGraSemOlaSoaLotVanMadZidMenBosGreMonSvaNolAinAkiAlfAndAttBarBauBehBeiBiaBinBloBocBorBosBotBreBriBurCarCarCarCasCasChrCimClaConCouCroDeDeDeDeaDelDirDjaDjaDruEldErsFriGenGeoGoeGonGraGraGreHanHasHayHelHenHerHinHofHunJohKhaKlaKopKreKuzLauLilLipLocLotLulMaaMacMadMarMasMenMeyMonMorMouMpoNikNolOlaOliPaaPelPerPflPitPooRafRenRisRoeRosRudSafSanSanSarScaSchSchSchSemShaSinSkaSkrSmySoaSvaSzcTacTagTeThiTjeTreTruTruTurUrlVaaVanVanWilWneWylYeuZid:2021:erc}.
Unfortunately, precordial thump and percussion pacing have not yet been sufficiently investigated and the American Heart Association (AHA) recommends against using these procedures in a general clinical setting since existing clinical studies on these techniques do not show strong evidence of being effective or ineffective~\cite{OlaManPerAviBroCasChuConCouEscHatHunKudLimNisRisSemSmiSmyVaiNolHazMorSvaRafKuzGraDeeSmiRaj:2020:abl}. 
Here, computational analysis may supply the necessary insight, provided that suitable models for an accurate description of the back-coupling are available. 
Interestingly, despite the importance of emergency pacing techniques, no theoretical framework exists to understand its mechanisms in detail. 

Understanding the mechanism linking pacing and mechanical stimuli is certainly difficult and there have not been many experimental studies in this area~\cite{IzuKohBoyMiuBanChiTraBerChe:2020:mmc}.
In this context, \citet{QuiKoh:2016:cmr} investigated a rabbit heart subjected to several sequences of mechanical and electrical stimuli during sinus rhythm. The rabbit heart is in Langendorff perfusion with an inserted intraventricular balloon to simulate blood pressure.
During the application of mechanical and electrical stimuli to the ventricular epicardium, the balloon pressure and transmembrane voltage at a point on the ventricular epicardium are recorded.
In these experiments, several key features are identified: 1) There is a reversible, frequency-dependent loss of capture of mechanical pacing that depends on the number of applied mechanical stimuli; 2)  Alternating the application of mechanical and electrical stimuli leads to a faster loss of capture (in the number of applied mechanical stimuli) than when only mechanical pacing is applied.
The coupling between mechanical and electrical pacing found in these experiments cannot be explained by current models and highlights the importance of developing a theoretical framework for mechanical pacing of cardiac tissue.

Stretch-activated ion channels modulate ionic balances in cells including cardiomyocytes 
{(we refer interested readers to the work of ~\citet{Sac:2010:sic} and ~\citet{PeyNerKoh:2016:cmi} for a detailed introduction to stretch-activated ion channels).}
In this paper, we assume that PIEZO channels are a major contributor to this mechanism. %
The PIEZO protein family consists of mechanically activated cation-selective channels~\cite{CosXiaSanSyeGraSpeKimSchMatDubMonPat:2012:ppa,LewGra:2021:pic}, which govern a wide range of physiological functions across several animal kingdoms~\cite{CosMatSchEarRanPetDubPat:2010:ppa}.
Vertebrates express primarily the isoforms Piezo1 and Piezo2~\cite{CosMatSchEarRanPetDubPat:2010:ppa}, which can be functionally identified as biological sensors for mechanical stimuli~\cite{LewCuiMcDGra:2017:trm}.
The characteristic feature of PIEZO ion channels is a significantly voltage-dependent~\cite{MorSerFleSanLew:2018:vgm} slow (Piezo1) and fast (Piezo2) inactivation kinetics~\cite{WuYouLewMarKalGra:2017:ima}, which has been demonstrated to allow a frequency-dependent response to periodic mechanical stimuli.
Recent research has uncovered that this class of ion channels is involved in many physiological and pathophysiological processes.
In tactile neurons, Piezo2 plays a key role in the touch sensation of rough surfaces~\cite{LewCuiMcDGra:2017:trm}.
Experiments also suggest that PIEZO channels act as pressure overload sensors in ventricular cardiomyocytes~\cite{Bac:2022:plm,YuGonKesGuoWuLiCheZhoIisKaiGraCoxFenMar:2022:pcm} with implications in cardiovascular development~\cite{RanQiuWooHurMurCahXuMatBanCosLiChiPat:2014:pma,LiHouTumMurBruLudSedHymMcKYouYulMajWilRodBaiKimFuCarBilImrAjuDeaCubKeaPraEvaAinBee:2014:piv} and homeostasis~\cite{LiHouTumMurBruLudSedHymMcKYouYulMajWilRodBaiKimFuCarBilImrAjuDeaCubKeaPraEvaAinBee:2014:piv,Bac:2022:plm,BarEvaBlySteChuDebMusLicParFutTurBee:2022:gpg,Lim:2022:psp,YuGonKesGuoWuLiCheZhoIisKaiGraCoxFenMar:2022:pcm,YuGonKesGuoWuLiIisKaiGraCoxFenMar:2022:ptw,CahLukRanChiBanPat:2015:plm}.
In addition, experiments have been able to link Piezo1 upregulation to pathophysiological processes such as  hypertrophy~\cite{ZhaSuLiMaSheWanSheCheJiXieMaXia:2021:pmp,Bac:2022:plm,JiaYinWuZhaWanCheZhoXia:2021:mpc}, arrhythmia~\cite{JiaYinWuZhaWanCheZhoXia:2021:mpc}, and heart failure~\cite{LiaHuaYuaCheLiaZenZheCaoGenZho:2017:scp,JiaYinWuZhaWanCheZhoXia:2021:mpc}.
Although it has been shown that PIEZO ion channels are expressed in the ventricular myocardium~\cite{LiaHuaYuaCheLiaZenZheCaoGenZho:2017:scp}, it has only been recently demonstrated that Piezo1 channels are located in the transverse tubular system of cardiomyocytes~\cite{JiaYinWuZhaWanCheZhoXia:2021:mpc,YuGonKesGuoWuLiCheZhoIisKaiGraCoxFenMar:2022:pcm}.
Interestingly Piezo1 and Piezo2 are involved in physiological processes outside the neural and cardiovascular systems, as mutations have been linked to severe diseases, e.g., hereditary xerocytosis~\cite{ZarSchHouMakHouSmiRinGal:2012:mmp,BaeGnaNicSacGot:2013:xcm}, Marden-Walker syndrome, and Gordon syndrome~\cite{McMBecChoShiBucGilAraAylBitCarCleCroCurDevEveFryGibGioGraHalHecHeiHurIraKraLerMowPlaRobSchScoSeaSheSplSteStuTemWeaWhiWilTabSmiSheNicBam:2014:mpc} in humans.

In this work, we will analyze a simplified version of the experiment by \citet{QuiKoh:2016:cmr} via mathematical modeling and simulation studies. 
As a first step, we will motivate why we believe that these experimental observations are related to Piezo1 proteins, {which are voltage-modulated~\cite{MorSerFleSanLew:2018:vgm} and stretch-activated~\cite{CosMatSchEarRanPetDubPat:2010:ppa,MulGhaLeeDubAarMarSpeReiHenChePat:2023:doc} ion channels -- this form of electromechanical gating is a unique and distinguishing feature among all known ion channels}.
Subsequently, we will develop a Piezo1 voltage-modulated, continuous-time Markov chain model on the basis of the experiments in~\cite{LewCuiMcDGra:2017:trm,LewGra:2021:pic} and~\citet{MorSerFleSanLew:2018:vgm}.
This ion channel model will then be integrated in the well-known rabbit ventricular cardiomyocyte model by~\citet{MahShiSatBahOlcXieYanCheResKarGarQuWei:2008:rva} to investigate the qualitative role of Piezo1~\cite{CosMatSchEarRanPetDubPat:2010:ppa} channels during mechanical and electromechanical pacing of the cardiac tissue.
Finally, we will discuss the model limitations and several possibilities to improve and refine the model in subsequent studies.

\section{Methods}

\subsection{Model Construction}

In this section we construct a mathematical model of a rabbit ventricular cardiomyocyte, which is capable of explaining, at least qualitatively, the experimental observations in~\citet{QuiKoh:2016:cmr}.
In these experiments, the authors applied electrical and mechanical stimuli to Langendorff-perfused rabbit hearts while measuring their electrical response.
Mechanical stimuli have been applied using a piston against the left ventricular (epicardial) wall, while electrical stimuli have been applied using an electrode to deliver local currents to the left ventricle epicardium.
Both mechanical and electrical stimulation sites have been chosen to be spatially close.
Another electrode on the left ventricular wall (epicardial) has measured the electrical activity.
The main observation is that, at first, the ventricle responds to the mechanical stimuli with a full depolarization, which we call \textit{mechanical capture}.
After a few mechanical stimuli (see, e.g., Fig.~\ref{fig:moroni-weak-rectification-experiment-reproduction}A), the ventricle stops responding, which is called \textit{loss of (mechanical) capture}.
The exact number of mechanical stimuli until loss of capture depends on the stimulus frequency and the number of applied electrical stimuli in-between two mechanical stimuli.
The observed loss of capture is reversible, and mechanical capture is restored after mechanical stimuli have not been applied for a sufficiently long time window~\cite{QuiKoh:2016:cmr,MorSerFleSanLew:2018:vgm}.

\subsubsection{Determination of the Molecular Candidate for the Ion Channel Model}
\label{sec:molecular-candidate}

The ion channels to be included in the model need to generate enough current to depolarize the membrane after sufficient {stretch}.
Generally, the currents generated by the {stretch of the} cells are categorized into two broad classes: non-selective stretch-activated ion channels $I_{\mathrm{SAC,NS}}$ and $\mathrm{K^+}$-selective stretch-activated ion channels $I_{\mathrm{SAC,K}}$ {for which several molecular candidates have been proposed~\cite{ReeKohPey:2014:mcc}}.
For healthy cardiomyocytes in a physiological environment it can be shown that, assuming an Ohmic current generated by the flux of ions through an ion channel, the reversal potential of $\mathrm{K^+}$ ($\approx$ \SI{-90}{\milli\volt}) is below the depolarization threshold ($\approx$ \SI{-60}{\milli\volt}).
Hence, activating these ion channels cannot induce membrane depolarization. 
We refer to~\cite{QuiJinLeeKoh:2017:mie} for more details.
This narrows down the candidates for the ion channels.
{The PIEZO family and the TRP family of ion channels are hypothesized to be primary candidates for $I_{\mathrm{SAC,NS}}$~\cite{ReeKohPey:2014:mcc}.}

Obviously, the candidate ion channels also have to be expressed in real cardiomyocytes.
In addition, we assume that the ion channel is located on the cell membrane to depolarize, since this location may be more responsive to mechanical {stretch}.
Although we acknowledge the possibility that the hypothetical ion channel can be located inside the sarcoplasmatic reticulum, we do not follow this idea in the first iteration of the integrated model proposed here.
This choice restricts the possible ion channel candidates, as some TRP channels are not located on the cell membrane (see, e.g.,~\cite{YueXieYuStoDuYue:2015:rtc} for an overview).
For the PIEZO family, it has been shown in experimental studies that both channels expressed in vertebrates are localized near the transverse tubular system of cardiomyocytes~\cite{JiaYinWuZhaWanCheZhoXia:2021:mpc,KloMeaWeiSteGeeStaLinSchReiEscHir:2022:pni}.
Note that additional experiments suggest the existence of a protective mechanism downregulating PIEZO channels' activity when these channels are integrated into the endoplasmatic membrane~\cite{ZhaChiJiaZhaXia:2017:pim}.
These experiments provide further evidence that Piezo1 channels play a significant role in the cardiomyocyte's electromechanical response, which is the focus of the current work. %

Finally, based on the experimental observation, we hypothesize that the ion channel needs to feature slow frequency-dependent inactivation.
This rules out the possibility that only TRP channels will be responsible for the behavior observed in mechanical pacing experiments by~\cite{QuiKoh:2016:cmr} and leaves the PIEZO family as an optimal candidate, as the experiments of~\cite{LewCuiMcDGra:2017:trm} provide strong evidence for their frequency-dependent behavior.
The PIEZO family has two known members, Piezo1 and Piezo2.
The latter features fast inactivation (see, e.g.,~\cite{LewCuiMcDGra:2017:trm}), which further narrows down our choice to the Piezo1 channel.

More evidence against the TRP channels being the primary candidate to explain the experimental observation in~\cite{QuiKoh:2016:cmr} is provided by recent studies.
Experiments by~\citet{NikCoxRidRohCorVasLavMar:2019:mti} suggest that most TRP channels might not be inherently stretch-activated.
Instead, they hypothesize that TRP channels play a role downstream in mechanosensation pathways.
This hypothesis is further supported by more recent experiments on the interplay between TRPM4 and Piezo1, showing colocalization of these ion channels in the transverse-axial tubular system of ventricular cardiomyocytes of different species~\cite{JiaYinWuZhaWanCheZhoXia:2021:mpc,YuGonKesGuoWuLiCheZhoIisKaiGraCoxFenMar:2022:pcm}.

\subsubsection{Improved Piezo1 Continuous-Time Markov Chain Model}

Three distinct mathematical models for Piezo1 have been proposed in the literature.
\citet{BaeGotSac:2013:hpr} have proposed a continuous-time Markov chain model with three states (open, closed, and inactivated), which are fully connected and which feature a pressure-dependent transition from the closed to the open state and from the inactivated to the closed state.
This is an extension of a linear Markov chain where the closed and inactivated states are disconnected, see~\cite{GotBaeSac:2012:gmc}.
This model has been shown to capture well the fast inactivation behavior in response to single mechanical stimuli.
Modal analysis of a coarse-grained Piezo1 model~\cite{ZheSac:2017:isd} further supports the model assumption that there are indeed at least three independent states.

\citet{LewCuiMcDGra:2017:trm} have proposed a similar model.
Based on the argument that the inactivation kinetics of PIEZO ion channels in two-step pressure stimulation protocols can be fitted well with two exponentials, \citet{LewCuiMcDGra:2017:trm} hypothesized that there are two distinct inactivation states.
They corroborated this hypothesis by showing that the three state models~\cite{GotBaeSac:2012:gmc,BaeGotSac:2013:hpr} fail to explain a new set of experiments on frequency-dependent stimulus response of mouse Piezo1 and Piezo2 channels.
Therein Piezo1 has been shown to act as a band-pass filter on sinusoidal pressure stimuli.
Their proposed model contains four states, containing the same cycle as~\citet{BaeGotSac:2013:hpr}'s model and an additional second inactivation state with constant rates, which has been bidirectionally connected to the open state.
\citet{LewCuiMcDGra:2017:trm} showed that this model can capture well the frequency-dependent behavior of Piezo1 and Piezo2 ion channels.
Although we were able to identify a recent experimental investigation of the mechanism of inactivation by~\citet{ZheGraBag:2019:hgi} supporting the hypothesis of the existence of two possibly distinct inactivation states for the Piezo1 channel, further studies are required to determine whether these states are coupled or truly distinct.
In addition, the model from \citet{LewCuiMcDGra:2017:trm} lacks the voltage-dependent behavior seen in other experiments~(e.g.,~\cite{WuYouLewMarKalGra:2017:ima,MorSerFleSanLew:2018:vgm}).

Finally, there is the recent Hodgkin-Huxley formulation by~\citet{ZhaZou:2022:ntm} with three independent gates featuring both voltage-modulation and stretch activation, which distinguishes it from the previous models.
A possible drawback of this model is that existing simulations on the molecular dynamics of Piezo1~\cite{ZheSac:2017:isd,DeBeeKal:2021:mds,ChoDeHymPovLudShiBeeKal:2021:mfp} and structural investigations of it~\cite{SaoMurKefWhiPatWar:2018:sma} suggest a global transition structure between independent energy states of the ion channel, as reflected in the previously published Markov chain models discussed in the paragraphs above.
Furthermore, experiments on the electrophysiological properties of Piezo1 suggest a voltage-dependence between inactivation and activation~\cite{WuYouLewMarKalGra:2017:ima,MorSerFleSanLew:2018:vgm}.
This violates the assumption of Hodgkin-Huxley models that the ion channel opens via independently acting gates.
The authors provide different sets of parameters per experiment and per applied stimulus in each experiment, resulting in no continuous dependence of the gating variables on the pressure, underlining this problem.
While this does not exclude the existence of a Hodgkin-Huxley type model, a different approach may avoid the outlined issue and therefore we will not consider Hodgkin-Huxley type models for the Piezo1 channel in this work.

With the information gathered from these previous models, we constructed a 4 state Markov chain shown in figure~\ref{fig:piezo1-ctmc-model}.
We started the model construction with the Markov chain described by~\citet{LewCuiMcDGra:2017:trm}.
We explored many different formulations, from linear to more complex topologies with multiple cycles.
However, we decided to present this topology, since it is connected to other numerical experiments, as a similar model has been used in previous computational studies of Piezo1, and since it yielded sufficiently good fits to the data with a comparably low number of parameters.
To introduce the voltage-modulation into the model, we added a linear voltage term in each exponent, and the resulting model retained, at least in principle, the frequency-dependent behavior described in~\citet{LewCuiMcDGra:2017:trm}.
{To preserve the bandpass-filtering ability of the original model by~\citet{LewCuiMcDGra:2017:trm}, at least in principle, we also keep the transition rates from the closed to the fast inactivating state, and from the fast inactivating state to the open state to be pressure-dependent}.
A description of the parameter optimization procedure can be found in Section~\ref{sec:optimization-strategy}.
Based on this procedure, we were unable to eliminate any of the voltage terms introduced in the model (through regularization with $\gamma > 0$, cf. eq.~\ref{eq:loss-function}).
\begin{figure}[t]
\centering
\begin{tikzpicture}[-{Latex[length=7pt,width=7pt]}, shorten >=2pt , line width = 0.5pt ,
node distance = 4cm]
    \node[circle, draw] (open) {$O$};
    \node[circle, draw] (closed) [above of=open] {$I_{1}$};
    \node[circle, draw] (inactive_fast) [left of=open] {$C$};
    \node[circle, draw] (inactive_slow) [right of=open] {$I_{2}$};

    \path (open) edge node[above,sloped] {$r_1 e^{c_1^{\mathrm{e}}\Delta\mu}$} (closed);
    \path (closed) edge[bend left] node[above,sloped] {$r_2 e^{c_2^{\mathrm{m}}p + c_2^{\mathrm{e}}\Delta\mu}$} (open);
    \path (closed) edge node[below,sloped] {$r_3 e^{c_3^{\mathrm{e}}\Delta\mu}$} (inactive_fast);
    \path (inactive_fast) edge[bend left] node [above,sloped] {$r_4 e^{c_4^{\mathrm{m}}p + c_4^{\mathrm{e}}\Delta\mu}$} (closed);
    \path (inactive_fast) edge node [above,sloped] {$r_5 e^{c_5^{\mathrm{e}}\Delta\mu}$} (open);
    \path (open) edge[bend left] node [below,sloped] {$r_6 e^{c_6^{\mathrm{e}}\Delta\mu}$} (inactive_fast);
    \path (inactive_slow) edge node [above,sloped] {$r_7 e^{c_7^{\mathrm{e}}\Delta\mu p}$} (open);
    \path (open) edge[bend right] node [below,sloped] {$r_8 e^{c_8^{\mathrm{e}}\Delta\mu}$} (inactive_slow);
\end {tikzpicture}
\caption{Continuous-time Markov chain of the proposed Piezo1 model extending the model of~\cite{LewCuiMcDGra:2017:trm}.
Analogously to the original work, the proposed model contains an open state $O$, fast and slow inactivation states $I_{1}$ and $I_{2}$, and a closed state $C$.
$p$ denotes the pressure and $\Delta\mu$ the electrochemical driving force acting on the channel.
Since the reversal potential of PIEZO channels is hypothesized to be approximately 0 (i.e., $E_r = 0$, see, e.g,~\cite{GotBaeSac:2012:gmc,CosMatSchEarRanPetDubPat:2010:ppa}), we obtain $\Delta\mu = \varphi_{\mathrm{m}} - E_r = \varphi_{\mathrm{m}}$, where $\varphi_{\mathrm{m}}$ denotes the transmembrane potential.
Exponential transition functions have been chosen in conformance to Eyring's classical transition state theory~\cite{Eyr:1935:aca,Eyr:1935:acc}.
The parameters are grouped into rate coefficients $r_i$, mechanical pressure coefficients $c^m_i$ and electrochemical driving force coefficients $c^e_i$.
We want to highlight that the transition rate between the slow inactivating state $I_{2}$ and the open state $O$ is made dependent on the product of the electrochemical driving force and the pressure.
\label{fig:piezo1-ctmc-model}}
\end{figure}
For the model construction procedure, we have utilized \textit{Catalyst.jl}~\cite{LomMaIliGowKorYewRacIsa:2022:cfb} and \textit{ModelingToolkit.jl}~\cite{MaGowAnaLauShaRac:2021:mcg}.
The model is also provided as supplementary material via \textit{CellML}~\cite{CleCooCooGarMoyNicNieSor:2020:c}, where it has been validated via \textit{OpenCOR}~\cite{GarHun:2015:Omi}.

\subsubsection{Ventricular Cardiomyocyte Model}
\label{sec:VentrCardioModel}

Our main goal is to derive a cell model able to describe as many mechanical pacing experiments by~\citet{QuiKoh:2016:cmr} as possible. 
These experiments were conducted on rabbit hearts at high pacing rates, an thus we decided to utilize the Mahajan-Shiferaw ventricular cardiomyocyte model~\cite{MahShiSatBahOlcXieYanCheResKarGarQuWei:2008:rva} as our foundation.
Indeed, the Mahajan-Shiferaw model was constructed for exactly this type of scenarios.
\begin{figure}[t]
    \centering
    \includegraphics[width=\textwidth]{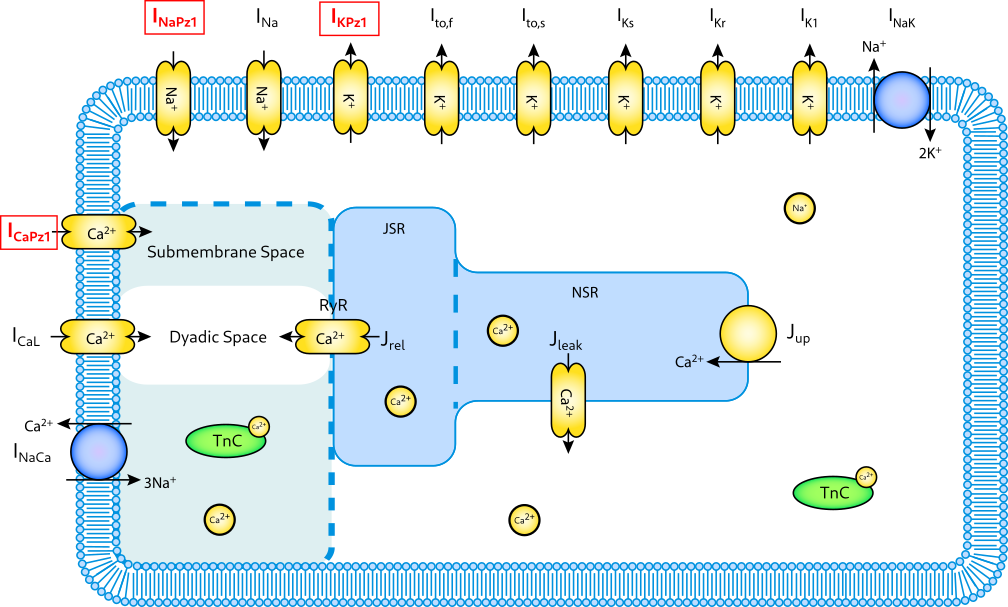}
    \caption{
        Schematic of the proposed lumped parameter cell model based on the Mahajan-Shiferaw rabbit ventricular cardiomyocyte model~\cite{MahShiSatBahOlcXieYanCheResKarGarQuWei:2008:rva}.
        The new Piezo1 associated currents ($I_{\mathrm{KPz1}}$, $I_{\mathrm{NaPz1}}$ and $I_{\mathrm{CaPz1}}$) are highlighted inside boxes with bold-red fonts.
        JSR is the junctional sarcoplasmatic reticulum and NSR is the non-junctional sarcoplasmatic reticulum.
        The remaining currents and fluxes are defined as in the original model~\cite{MahShiSatBahOlcXieYanCheResKarGarQuWei:2008:rva}.
    \label{fig:modified-mahajan-model}
    }
\end{figure}
We have integrated the new Piezo1 model as follows under some assumptions regarding the induced currents.
First, we assume that the ions do not interact with each other in the ion channel.
Second, the generated current is Ohmic.
{Third, we assume that the ion channel is non-selective and hence conducts $\mathrm{K}^+$, $\mathrm{Na}^+$ and $\mathrm{Ca}^{2+}$, as suggested by a range of experiments, e.g., \cite{CosMatSchEarRanPetDubPat:2010:ppa,CosXiaSanSyeGraSpeKimSchMatDubMonPat:2012:ppa,GnaBaeGotSac:2015:isp,Got:2017:tf}}.
This translates to the following current formulation for ions flowing through the Piezo1 channel
\begin{equation}
    I_{\mathrm{Pz1}}
    = \underbrace{p_{\mathrm{O}} \overline{g}_{\mathrm{K}}(\varphi_{\mathrm{m}} - E_{\mathrm{K,s}})}_{:=I_{\mathrm{KPz1}}}
    + \underbrace{p_{\mathrm{O}} \overline{g}_{\mathrm{Na}}(\varphi_{\mathrm{m}} - E_{\mathrm{Na,s}})}_{:=I_{\mathrm{NaPz1}}}
    + \underbrace{p_{\mathrm{O}} \overline{g}_{\mathrm{Ca}}(\varphi_{\mathrm{m}} - E_{\mathrm{Ca,s}})}_{:=I_{\mathrm{CaPz1}}}
    \, ,
    \label{eq:IPz1}
\end{equation}
where $p_{\mathrm{O}}$ is the open probability of the Piezo1 channel, given by the probability of being in the open state $O$ of the novel Markov chain formulation shown in fig.~\ref{fig:piezo1-ctmc-model}. 
The quantities $\overline{g}_{\mathrm{K}}$, $\overline{g}_{\mathrm{Na}}$, and $\overline{g}_{\mathrm{Ca}}$ are the maximal normalized conductances for the respective ion fluxes, and $E_{\mathrm{K,s}}, E_{\mathrm{Na,s}}, E_{\mathrm{Ca,s}}$ are the corresponding Nernst potentials (cf.~\cite[Ch.3.1]{KeeSne:2009:MPC})
related to the submembrane space ion concentrations.
Since the Mahajan cell model only tracks intracellular concentrations for $\mathrm{Na^+}$ and assumes constant intracellular $\mathrm{K^+}$, we approximate $E_{\mathrm{K,s}}$ and $E_{\mathrm{Na,s}}$ as in the original model via
\begin{equation}
    E_{\mathrm{Na,s}} \approx \frac{R T}{F}\ln \frac{\left[\mathrm{Na^+}\right]_{\mathrm{o}}}{\left[\mathrm{Na^+}\right]_{\mathrm{i}}} \, , \quad
    E_{\mathrm{K,s}} \approx \frac{R T}{F}\ln \frac{\left[\mathrm{K^+}\right]_{\mathrm{o}}+pr_{\mathrm{NaK}}\left[\mathrm{Na^+}\right]_{\mathrm{o}}}{\left[\mathrm{K^+}\right]_{\mathrm{i}}+pr_{\mathrm{NaK}}\left[\mathrm{Na^+}\right]_{\mathrm{i}}} \, , 
\end{equation}
where $pr_{\mathrm{NaK}}$ is unknown and will be studied in the results section.
The Nernst potentials for $E_{\mathrm{Ca,s}}$ are computed as
\begin{equation}
    E_{\mathrm{Ca,s}} = \frac{R T}{2 F}\ln \frac{\left[\mathrm{Ca^{2+}}\right]_{\mathrm{o}}}{\left[\mathrm{Ca^{2+}}\right]_{\mathrm{s}}} \, .
\end{equation}
The resulting current $I_{\mathrm{Pz1}}$ is added to the total transmembrane current.

In the next step we simplify the expressions for the maximal conductances by incorporating information from experiments.
We first assume that the conductances of Piezo1 between species are approximately equal.
\citet{GnaBaeGotSac:2015:isp} 
found
that, for single Piezo1 ion channels in isolation, the conductance of $\mathrm{Ca^{2+}}$ is $\approx 12$ pS.
In the absence of $\mathrm{Ca^{2+}}$, the conductance of $\mathrm{Na^{+}}$ is about $80\%$ the conductance of $\mathrm{K^{+}}$.
Together with the information that the reversal potential of Piezo1 is about \SI{0}{\milli\volt}, we obtain a conductance of $g_K = g_{Ca} E_{Ca,s}/(-E_{K,s}-0.8E_{Na,s}) \approx 52 \,\si{\pico\siemens}$ for $\mathrm{K^{+}}$, which is close to the reported range $47-53 \, \si{\pico\siemens}$ between \SI{-80}{\milli\volt} and \SI{-100}{\milli\volt}~\cite{GnaBaeGotSac:2015:isp}.
\citet{GnaBaeGotSac:2015:isp} have also shown that the conductance of $\mathrm{K^{+}}$ is reduced by about $25\%$ in the presence of $Ca^{2+}$.
With the additional information that for Ohmic ion channels we have $\overline{g}_i := N_{Pz1} g_i$, this allows us to reduce the number of parameters, as now only the average number of Piezo1 ion channels per cell ($N_{Pz1}$) is unknown.
Therefore, we study the effect of $N_{Pz1}$ in the results section as a normalized scaling parameter $scaling := N_{Pz1}10^{-6}/52$.

The new currents now change the ionic fluxes, which have to be added to the model.
In the following $J_{\square}$ denotes an ionic flux and $I_{\square}$ the associated ionic current, as defined in eq.~\eqref{eq:IPz1}.
Experimental studies suggest that Piezo1 channels are primarily located in the t-tubular membrane~\cite{JiaYinWuZhaWanCheZhoXia:2021:mpc,KloMeaWeiSteGeeStaLinSchReiEscHir:2022:pni}.
Since we hypothesize that the influx of ions has no significant \textit{direct} contribution to the calcium cycling within the dyadic clefts -- and hence to calcium-induced calcium release (CICR) -- in our model we do not link the Piezo1 channel directly to the dyadic clefts but rather to the submembrane space. 
In this location, the Piezo1 channel indirectly contributes to the calcium cycling via membrane depolarization and the release of calcium to the membrane subspace near the dyadic clefts, as highlighted in the schematic cell model in Fig.~\ref{fig:modified-mahajan-model}.
To enforce conservation of the ionic concentrations consistently with the remaining Mahajan-Shiferaw model, we have to modify the evolution equations for the submembrane calcium concentration $\left[\mathrm{Ca^{2+}}\right]_{\mathrm{s}}$ 
\begin{equation}
    \mathrm{d}_t [\mathrm{Ca^{2+}}]_{\mathrm{s}} = \beta_{\mathrm{s}} \left( \frac{v_{\mathrm{i}}}{v_{\mathrm{s}}} (J_{\mathrm{rel}} - J_{\mathrm{d}} + J_{\mathrm{CaL}} + \boldsymbol{\color{red}J_{\mathrm{CaPz1}}} + J_{\mathrm{NaCa}}) - J^{\mathrm{s}}_{\mathrm{trpn}} \right) \, ,
\end{equation}
where the flux is consistently computed as $J_{\mathrm{CaPz1}} = -\frac{C_{\mathrm{m}}}{2 F v_{\mathrm{i}}} I_{\mathrm{CaPz1}}$, and the evolution of the intracellular sodium concentration $[\mathrm{Na^{+}}]_{\mathrm{i}}$
\begin{equation}
    \mathrm{d}_t [\mathrm{Na^{+}}]_{\mathrm{i}} = \alpha' \left( I_{\mathrm{Na}} + \boldsymbol{\color{red}I_{\mathrm{NaPz1}}} + 3 I_{\mathrm{NaCa}} + 3 I_{\mathrm{NaK}} \right) \, ,
\end{equation}
to include the novel ionic flux.
Note that potassium is not explicitly tracked in this model.
Here $\alpha'$ is a factor to translate the Ohmic currents to its corresponding ion flux with correct magnitude, $\beta_{\mathrm{s}}$ a binding coefficient, $v_{\mathrm{i}}$ and $v_{\mathrm{s}}$ the representative volumes of the intracellular and submembrane compartments.
All values for these parameters and the formulations for the currents, fluxes, and their evolution laws, even the ones not directly presented here, are taken from the original paper~\cite{MahShiSatBahOlcXieYanCheResKarGarQuWei:2008:rva}.
A schematic of the proposed lumped parameter cell model is given in Fig.~\ref{fig:modified-mahajan-model}.
The modified cell model, together with scripts for all experiments, are provided in the supplementary materials.
Our novel ion channel model is integrated into the Mahajan-Shiferaw via \textit{CellML 2.0}~\cite{CleCooCooGarMoyNicNieSor:2020:c} using \textit{OpenCOR}~\cite{GarHun:2015:Omi}.

\subsection{Parameter Optimization}
\label{sec:optimization-strategy}

The parameters for the new Markov chain model have been obtained by formulating an optimization problem to fit some of the transmembrane potential trajectories from the experiments presented in~\cite{MorSerFleSanLew:2018:vgm} on outside-out patches of N2a cells containing wild type mouse Piezo1 channels.
All simulations required for the parameter adjustment have been performed using ABDF2~\cite{CelAguCha:2014:iab} and \textit{Sundials CVODE}~\cite{HinBroGraLeeSerShuWoo:2005:ssn} (for validation) via DifferentialEquations.jl~\cite{RacNie:2017:djp}.
These schemes provided good trade-offs between robustness and performance in the simulations.
To find a suitable set of parameters, we have decided to minimize a normalized Huber-type Lasso loss function with normalized data:
\begin{equation}
    L(\bm{p},\bar{\bm{g}}) = \sum_{E \in \mathcal{E}} \sqrt{\frac{1}{\abs{T_E}} \sum_{t_i \in T_E} H\left(P_O(t_i;\bm{p})g_i\frac{\varphi_{\mathrm{m}_E}}{\varphi_{\mathrm{m}_{\max}}} - \bar{i}_E(t_i)\right)} + \gamma ||\bm{p}||_1,
    \label{eq:loss-function}
\end{equation}
where $H$ is the Huber function~\cite{Hub:1964:rel} with continuation at $1$, i.e. $H(x) = \min(\abs{x}, x^2)$, $\mathcal{E}$ is the set of experiments, $T_E$ is the set of time points at which the normalized currents $\bar{i}_E(t_i)$ are known, $\varphi_{\mathrm{m}_E}$ is the clamped voltage of the current experiment in \si{\milli\volt}, $\varphi_{\mathrm{m}_{\max}} = \SI{80}{\milli\volt}$ is the absolute value of the maximum clamped voltage, and $P_O(t_i;\bm{p})$ is the open probability of a simulated group of Piezo1 ion channels at a specific time point $t_i$ using the parameters $\bm{p}$.
Note that, while the experiments~\cite[Fig. 1b]{MorSerFleSanLew:2018:vgm} revealed that the conductivity of Piezo1 is slightly nonlinear, it can be approximated reasonably well by a constant.
Furthermore, we use different bulk conductivities per experiment, denoted by $g_i$ for experiment $i$, because the experiments have been conducted on different patches having different numbers of ion channels, resulting in different observed currents.

As the parameter optimization problem contains exponential functions, the chosen Huber-type loss has shown practical advantage over the classical root mean squared error (RMSE) in our experiments.
Gradient information far from the optimum are naturally more limited, since the derivative of the Huber function is naturally bounded through its linear tails.
The loss function (as well as other commonly used loss functions like RMSE) is not convex though. 
This is also partly due to the nature of the model itself, which contains parameters that appear nonlinearly in the equations. 
Hence, we have to deploy a global optimization procedure to obtain accurate parameters.
For the optimization we utilize L-BFGS~\cite{LiuNoc:1989:lmb} with random initial guesses (N=1000).
Initial parameter guesses were uniformly sampled from the interval $[-1,1]$ for each parameter.
For the parameter optimization, we have normalized the transition functions by choosing $p = \overline{p}/70$ and $\Delta\mu = \Delta\overline{\mu}/140$, such that $p \in [0,1]$ and $\Delta\mu \in [-1,1]$.

{Dwell-time analyses of Piezo1 channels suggest that detailed balance should hold for Piezo1 models, at least in voltage-clamp protocols~\cite{WijOzkLac:2022:yef}.
We assume that detailed balance will also hold for arbitrary electrochemical gradients in a physiologically reasonable range.
We enforce this condition by constraining the parameters in the reaction triangle, where
\[
    r_1 e^{c_1^{\mathrm{e}}\Delta\mu} r_3 e^{c_3^{\mathrm{e}}\Delta\mu} r_5 e^{c_5^{\mathrm{e}}\Delta\mu} = r_2 e^{c_2^{\mathrm{m}}p + c_2^{\mathrm{e}}\Delta\mu} r_4 e^{c_4^{\mathrm{m}}p + c_4^{\mathrm{e}}\Delta\mu} r_6 e^{c_6^{\mathrm{e}}\Delta\mu}
\]
needs to hold.
Henceforth, we obtain
\[
    r_2 = \frac{r_1 r_3 r_5}{r_4 r_6},
\]
\[
    c_2^{\mathrm{m}} = -c_4^{\mathrm{m}},
\]
\[
    c_2^{\mathrm{e}} = c_1^{\mathrm{e}}+c_3^{\mathrm{e}}+c_5^{\mathrm{e}}-(c_4^{\mathrm{e}}+c_6^{\mathrm{e}})
\]
as constraints on the parameters.
}

For each numerical experiment evaluated in the loss function, we first allow the model to equilibrate by simulating \SI{20}{\second} without mechanical stimuli (zero pressure stimulus) and a constant experiment-dependent voltage.
At the beginning of each numerical experiment, we set the initial probability to be concentrated at the state $C$, i.e., $C = 1$, and the other states are set to zero, i.e., $I_1 = I_2 = O = 0$.
This initialization procedure resembles the initial state of the real experimental setup. 
We proceed by simulating the protocols described in each experiment as shown in Figures 1a, 2a, 2c and 2f from~\cite{MorSerFleSanLew:2018:vgm}.
The experimental results shown in the remaining figures were reserved for validating the calibrated model. 
The optimized parameter set is given in table~\ref{tab:optimal-piezo-params}.
Our optimization studies yield that $\gamma=0$ leads to the best match with the experimental results, which suggests that all parameters are necessary to fit the experiments, i.e., all connections must be voltage-dependent in the proposed Markov chain model formulation.

\begin{table}[]
\begin{tiny}
    \begin{center}
    \begin{tabular}{|c|c|c|c|c|c|c|c|c|c|}
        \hline
        $c^{\textrm{e}}_1$ & $c^{\textrm{m}}_2$ & $c^{\textrm{e}}_2$ & $c^{\textrm{e}}_3$ & $c^{\textrm{m}}_4$ & $c^{\textrm{e}}_4$ & $c^{\textrm{e}}_5$ & $c^{\textrm{e}}_6$ & $c^{\textrm{e}}_7$ & $c^{\textrm{e}}_8$ \\
        \hline
        -2.3753126 & 12.101605 & -1.0013516  & -5.6010337 & -12.101605 & 0.70629007 & 6.46389 & -1.2173947 & 8.739162 & -3.364573 \\
        \hline
    \end{tabular}\par
    \begin{tabular}{|c|c|c|c|c|c|c|c|}
        \hline
        $r_1$ & $r_2$ & $r_3$ & $r_4$ & $r_5$ & $r_6$ & $r_7$ & $r_8$ \\
        \hline
0.02634342 & 0.00574267 & 0.0008000712 & -0.008945307 & -1.3529515e-5 & -6.206403e-5 & -6.603723e-6 & -0.00079357607 \\
        \hline
    \end{tabular}
    \end{center}
\end{tiny}
    \caption{
        Parameters of the Piezo1 model used in the computational studies presented in sec.~\ref{sec:results}.
    }
    \label{tab:optimal-piezo-params}
\end{table}

\section{Results}
\label{sec:results}

\begin{figure}[t]
\centering
\includegraphics[width=\textwidth]{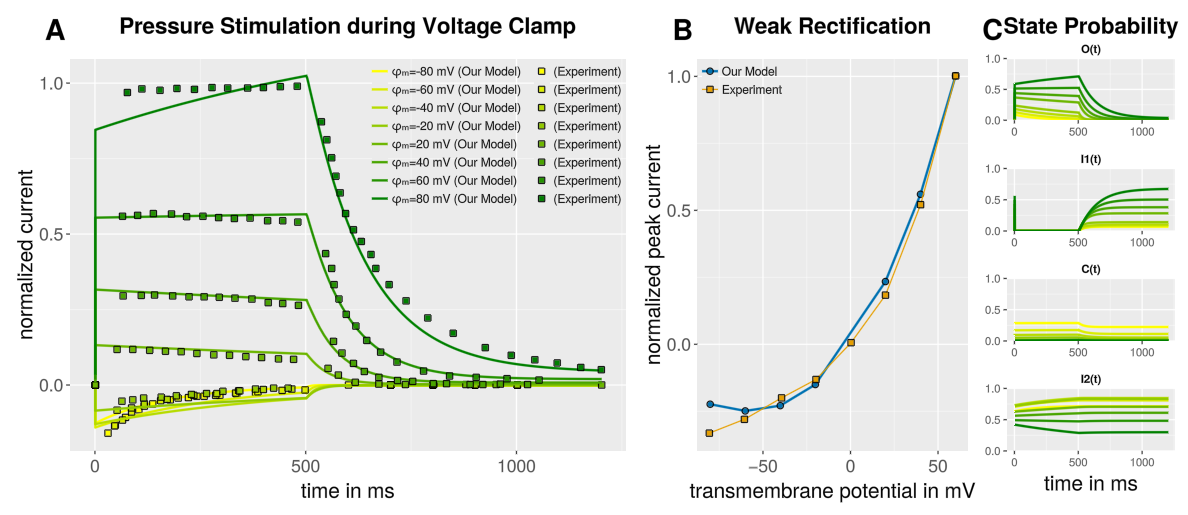}
\caption{Simulation {of the Piezo1 Markov chain model} vs. experimental results from~\cite[Fig.~1]{MorSerFleSanLew:2018:vgm} in the physiological regime of cardiomyocytes.
\textbf{A:} Normalized currents when applying a saturating pressure stimulus (length \SI{500}{\milli\second}) during voltage clamp (see legend $\varphi_{\mathrm{m}}$).
The predicted current traces match well the experimental results, although further quantitative improvements are possible.
The monoexponential, positively voltage-dependent current decay after releasing the pressure is consistent with the observation in~\cite{WuYouLewMarKalGra:2017:ima}.
\textbf{B:} Normalized current amplitudes during a pressure step in voltage clamp as described in~\cite{MorSerFleSanLew:2018:vgm}.
Our model also predicts qualitatively well the weak outward rectification of the ion channel.
\textbf{C:} Probabilities of the specific states corresponding to the simulations in \textbf{A}.
\label{fig:moroni-weak-rectification-experiment-reproduction}
}
\end{figure}

In this section we study the response of the modified Mahajan-Shiferaw cell model with integrated Piezo1 ion channel and how it can replicate published experimental results. 
In the first two studies, the Piezo1 model is initialized by setting the initial state $C = 1$ and $I_1 = I_2 = O = 0$.
The Piezo1 model is then simulated for 20 seconds using zero pressure and a constant voltage matching the voltage at the beginning of the corresponding computational experiment.\\
First, we investigate whether the voltage-modulated kinetics are sufficient to explain the voltage-dependent inactivation experiments shown in~\cite[Fig.~1]{MorSerFleSanLew:2018:vgm} and whether the ion channel shows weak rectification in the physiological voltage range. 
The results of this first study are summarized in Fig.~\ref{fig:moroni-weak-rectification-experiment-reproduction}. 
We observe that our proposed Markov chain captures well the voltage-dependent inactivation and predicts reasonably well the weak rectification behavior in the physiological voltage range for cardiomyocytes (\SI{-80}{\milli\volt} to \SI{50}{\milli\volt}).
The proposed model is also able to qualitatively represent the transient response during pressure-clamp experiments at different voltages.
When comparing to the data shown in~\cite[Fig.~1]{WuYouLewMarKalGra:2017:ima}, where a different experimental setup (i.e., with regard to cell type and stimuli protocol) is considered, the proposed model still represents well the experimental observations. 
Specifically, the decay after releasing the pressure stimuli is well represented, at least qualitatively, by our model even when the model parameters are not adjusted to this new experimental setup.

\begin{figure}[t]
\centering
\includegraphics[width=\textwidth]{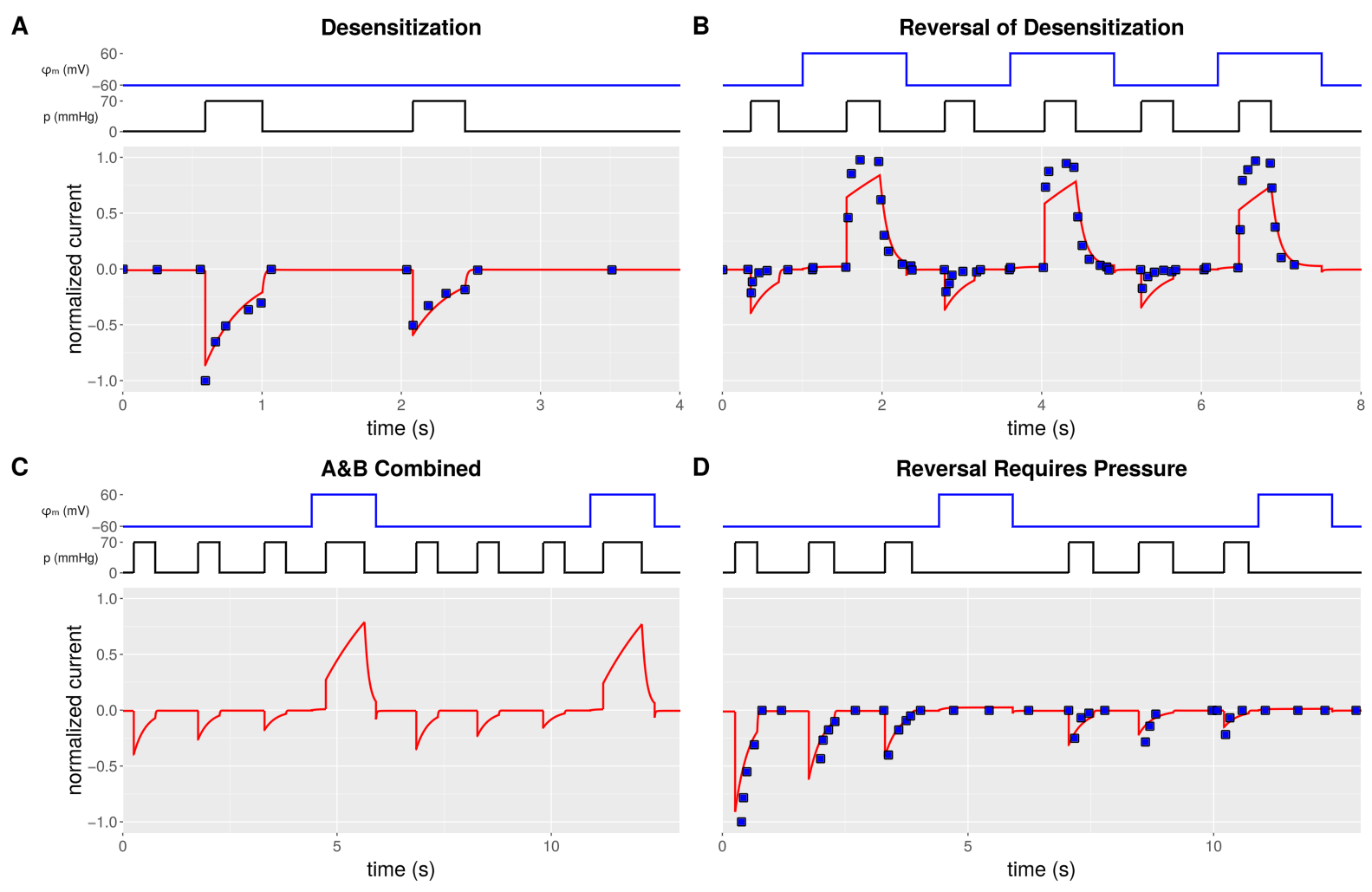}
\caption{
Simulation {of the Piezo1 Markov chain model} vs. experimental results from~\cite[Fig.~2]{MorSerFleSanLew:2018:vgm}.
Experimental data is shown using blue square markers and the corresponding numerical results are reported with a solid red line.
The curves above each subfigure describe the protocol in terms of applied pressure and transmembrane potentials over time.
\textbf{A:} PIEZO ion channels are known to desensitize when exposed to repeated mechanical stimulation at negative transmembrane potentials. Our model is able to reproduce this phenomenon well.
\textbf{B:} As shown in~\cite{MorSerFleSanLew:2018:vgm}, mechanical stimulation at positive transmembrane potential reverts desensitization.
\textbf{C\&D:} Reversal of desensitization requires positive electrochemical driving forces $\Delta\mu$ and pressure acting on the membrane as shown in~\cite{MorSerFleSanLew:2018:vgm}.
\label{fig:moroni-reset-experiment-reproduction}
}
\end{figure}

In a second set of computational studies of the Piezo1 model, its desensitization and reversibility is investigated~\cite{MorSerFleSanLew:2018:vgm}.
{Desensitization describes the phenomenon of an ion channel becoming less responsive to repeated stimuli of the same strength.}
To carry out these studies, we have designed protocols matching the described experiments in~\cite[Fig.~2]{MorSerFleSanLew:2018:vgm}.
As before, we compare the current traces computed by our model with the experimentally measured counterparts. 
The results are illustrated in Fig.~\ref{fig:moroni-reset-experiment-reproduction}. 
We can observe that key features are qualitatively well captured by our model.
The model desensitizes when exposed to pressure trains at negative electrochemical driving forces.
This desensitization is reset when a pressure step is applied at positive electrochemical driving forces.
When only exposed to positive electrochemical driving forces without applying a pressure step, the model does not significantly reset, as shown in the experiments.

\begin{figure}[h]
\centering
\includegraphics[width=\textwidth]{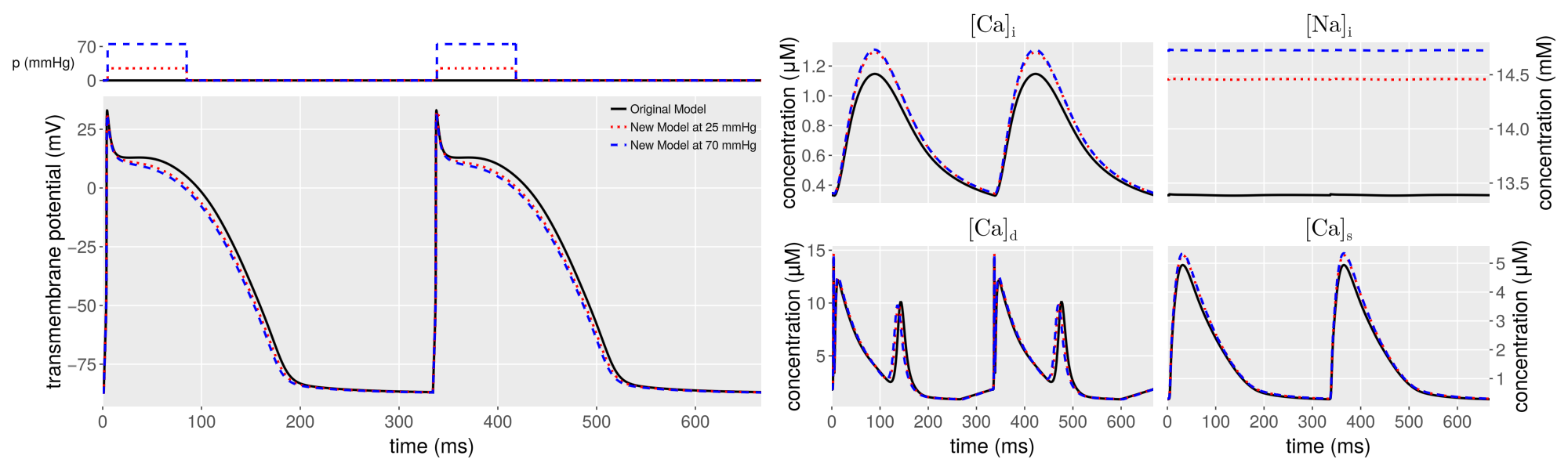}
\caption{
{Idealized pacing of the ventricular cardiomyocyte model at 3 Hz using the original cell model and the cell model with the addition of the proposed Piezo 1 ion channel. The Ca concentrations in different compartments (i, d, and s refer, respectively, to the intracellular, dyadic, and submembrane spaces), the Na concentration, and the action potential are investigated over the cardiac cycle. 
The original cell model provides a baseline that is compared with the results obtained using the cell models including the Piezo1 ion channel and stimulated with an additional rectangular pressure pulse of duration equal to \SI{80}{\milli\second} and magnitude p=\SI{25}{\mmHg} and p=\SI{70}{\mmHg}.
The pressure signal acts as a crude approximation of the stretch due to passive filling.
In these numerical experiments, we set $scaling=0.23$ and $pr_{\mathrm{NaK}}=0.067$. All models are paced 1000 times to reach the limit cycle, before results are collected for the next two cycles.}
\label{fig:cell-cycle-pacing}
}
\end{figure}

{
Next, we study the influence of the new ion channel on the transmembrane potential and ionic concentrations during an idealized cardiac cycle.
The updated rabbit ventricular cardiomyocyte model is periodically activated with a rectangular electrical stimulus of \SI{-15}{\nano\ampere\per\micro\farad} at a frequency of 3 Hz with width of \SI{3}{\milli\second}.
To model the mechanical stimulus, we start by dividing the cardiac cycle into two distinct mechanical regions, in which we assume that the cell is either compressed or stretched (in tension).
These regions roughly correspond to systole and diastole, although they are slightly shifted in time because cardiomyocyte shortening is not an instantaneous but a progressive process.
Due to this slight delay during the contraction phase of the ventricles, there is a brief period when the cardiomyocytes begin to increase their generated force, but are still (axially) stretched due to the the previous filling phase. %
Here we ignore the membrane stretch in the radial direction of contracted cardiomyocytes and neglect the complex cell geometry. According to this setup, an important assumption made in this model construction is that Piezo1 ion channels are only activated when the cardiomyocyte is axially stretched.
This common assumption is taken in most computational studies of stretch-activated ion channels~\cite{WalGucRatSun:2012:efr,ColPavSca:2017:emf,GerLoe:2024:dem} with some exceptions~\cite{LeeCanKal:2022:cmm}.
Furthermore, we need to relate the cardiomyocyte stretch with the pressure in the Piezo1 model.
Assuming a linear relationship between pressure and cardiomyocyte stretch where a 30\% stretch corresponds to the experimental pressure of \SI{70}{\mmHg} (p in the presented model) and that the representative stretch in diastole is about 10\%, we can identify a characteristic pressure of approximately \SI{25}{\mmHg} corresponding to this state.
This characteristic pressure of \SI{25}{\mmHg} (and \SI{70}{\mmHg}) is applied periodically at a frequency of 3 Hz.
The mechanical stimulus width is \SI{80}{\milli\second} and has a relative offset to the electrical stimulus of \SI{5}{\milli\second}.
The cell is prepaced for 1000 cycles to reach its limit cycle.
The 1001 and 1002 cycles are shown in Figure~\ref{fig:cell-cycle-pacing}.
}

{
Based on these studies, we want to highlight three interesting observations.
First, the intracellular calcium and sodium concentrations are elevated by approximately 8-10\% with respect to the original cell model without the Piezo 1 ion channel.
Second, the action potential is depressed and the action potential duration (APD) is shortened.
The latter is in line with the experimental observation that exposure of cardiomyocytes to YODA1, a Piezo1 agonist~\cite{SyeXuDubCosMatHuyMatLaoTulEngPetSchMonBanPat:2015:cam}, shortens the action potential~\cite{SuZhaLiXiLuSheMaWanSheXieMaXieXia:2023:cpe}.
Third, against our expectations, the application of stronger pressure stimuli does not result in a significant shift of intracellular calcium concentrations, but rather in a further elevation of intracellular sodium concentrations (Fig.~\ref{fig:cell-cycle-pacing}).
}

Finally, we study the experiments by~\citet{QuiKoh:2016:cmr} with our modified ventricular cardiomyocyte model.
{In these studies we approximate the intraventricular pressure of the experiment, as we were unable to design characteristic pressure profiles from the information provided in the original experimental paper~\cite{QuiKoh:2016:cmr}.}
We recall that, in the experiment by~\citet{QuiKoh:2016:cmr}, a Langendorff perfused rabbit heart is stimulated mechanically and electrically with different protocols.
The stimuli are applied with a linear piston and an electrode, respectively.
The key observations in this experiment are that: 1) mechanical capture is lost and the number of mechanical stimuli until loss of capture is inversely proportional to the stimuli frequency; and 2) loss of capture is faster (in terms of number of applied mechanical stimuli until loss of capture) when alternating mechanical and electrical stimuli are applied.
In order to systematically study the free parameters $pr_{\mathrm{NaK}}$ and $scaling$ (see sec.~\ref{sec:VentrCardioModel}), we introduce the simplifying assumption that we can study the qualitative response of the perfused heart at a material point level.
In making this simplifying assumption, we neglect all heterogeneities in cell types in the perfused heart, as well as all gradients in mechanical and electrical stimuli.
We discuss the limitations introduced by this simplification in detail in Section~\ref{sec:limitations}.\\

\begin{figure}[t]
\centering
\includegraphics[width=\textwidth]{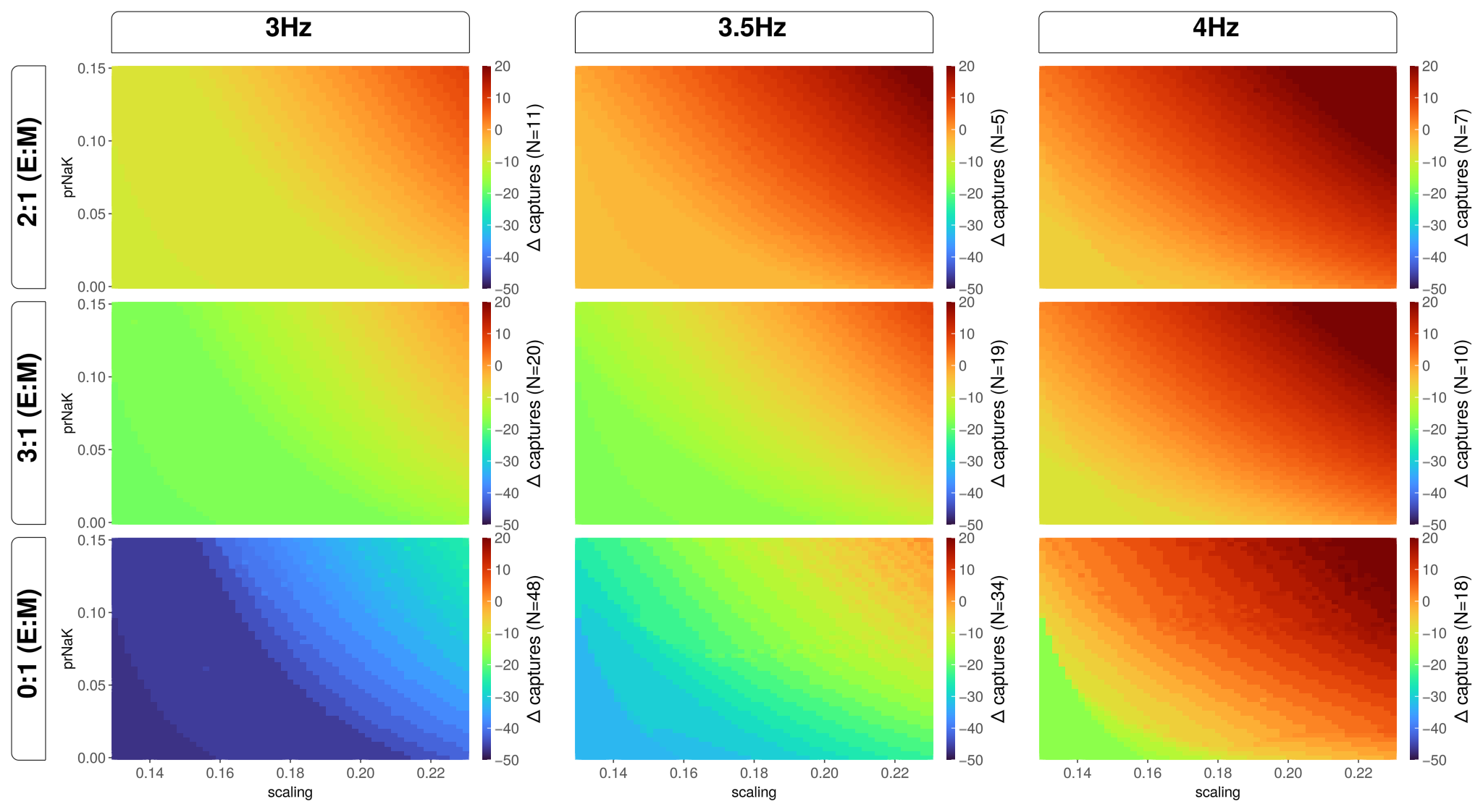}
\caption{
Differences in the total number of captured mechanical stimuli between the computational results and the experimental observations for different choices of $pr_{\mathrm{NaK}}$ and $scaling$ parameters in the proposed model (results are reported as number of captured mechanical stimuli in the simulations minus the number of captured mechanical stimuli in the experiments -- The latter is equal to the number `N' reported in each subfigure).
The scaling parameter corresponds to approximately 2,500 to 4,230 Piezo1 ion channels on the cell membrane.
Each plot represents a different experimental protocol (rows) and a different electromechanical pacing rate (columns).
In the associated colorbars, $N$ refers to the number of captured mechanical stimuli in the real experiment by~\citet{QuiKoh:2016:cmr} used as reference.
We can clearly observe a non-trivial relationship between the parameters and the number of captured mechanical stimuli in each experiment.
\label{fig:quinn-experiment-reproduction}}
\end{figure}

\begin{figure}[t]
\centering
\includegraphics[width=0.49\textwidth]{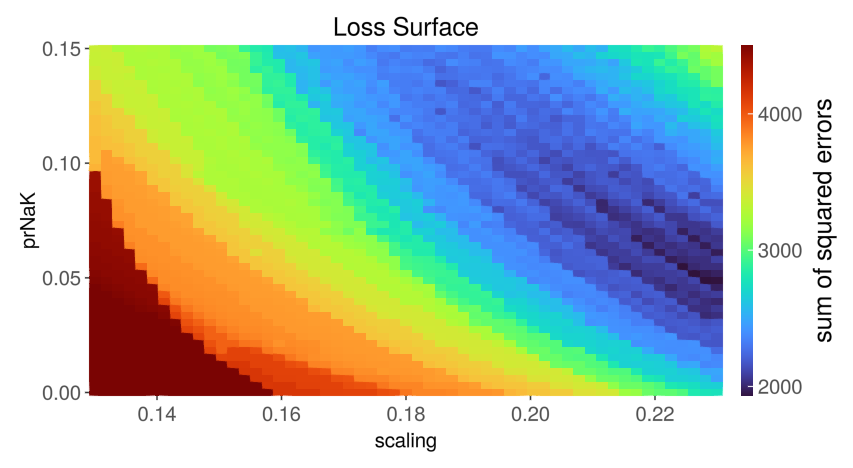}
\includegraphics[width=0.49\textwidth]{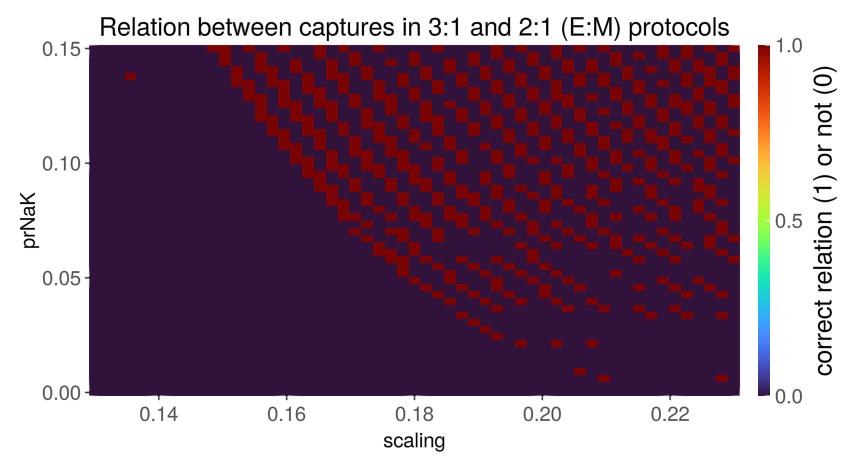}
\caption{
\textbf{Left:} {Surface of the loss function constructed as the} sum of the squared differences from Fig.~\ref{fig:quinn-experiment-reproduction}
across the different pacing protocol and frequencies showing the parameter values for $pr_{\mathrm{NaK}}$ and $scaling$ corresponding to the lowest overall error (see colorbar). 
\textbf{Right:} In the electromechanical pacing experiments by~\citet[Fig.~6]{QuiKoh:2016:cmr} it can be observed that the number of captured mechanical stimuli in the 3:1 (E:M) protocol is larger than the number of captured mechanical stimuli in the 2:1 (E:M) protocol.
All parameter combinations which represent this qualitative relation are shown as red pixels (see colorbar).
\label{fig:quinn-experiment-loss}}
\end{figure}

For our computational studies at the material point level, we approximate the experimentally applied mechanical stimuli (induced by a linear piston) as pressure steps with amplitude of $\SI{50}{\mmHg}$ for $\SI{10}{\milli\second}$.
The electrical stimulus is also modeled as a step function with an amplitude of $\SI{15}{\pico\ampere}$ and a width of $\SI{3}{\milli\second}$.
We explored the parameter space in two steps.
First, a coarse parameter grid was utilized (not shown) to narrow down the parameter space to reproduce the total number of captured mechanical stimuli recorded in the experiments.
In this first sweep of the parameter space, we identified that the conductance scaling parameter should be between 0.13 and 0.23, while the relative NaK contribution should be between 0.0 and 0.16.
This region was then analyzed using a fine grid.
The resulting difference in the total number of captured mechanical stimuli between the computational studies and experiment is presented in Fig.~\ref{fig:quinn-experiment-reproduction} for three protocols at three different frequencies.
The proposed model is able to reproduce, for some parameterizations, the experimental observation that the loss of capture in 2:1 (E:M) pacing occurs faster than in 3:1 (E:M) pacing.
However, none of the studied parameter combinations can reproduce all experimental findings by~\cite{QuiKoh:2016:cmr} quantitatively.
Furthermore, no simulation in the studied parameter range can reproduce the experimental observation that alternating mechanical and electrical stimuli lead to faster loss of mechanical capture than during mechanical pacing alone.
If we describe the study above as an optimization problem with standard sum of squared errors between the computational results and the experimental observations, then the resulting loss surface is presented in Fig.~\ref{fig:quinn-experiment-loss}. 

\section{Discussion \& Conclusion}

We have presented an improved Piezo1 ion channel model formulated as a continuous time Markov chain and integrated it into a {well-established} ventricular cardiomyocyte model~\cite{MahShiSatBahOlcXieYanCheResKarGarQuWei:2008:rva}.
To the best of our knowledge, this is the first time that a {Markov chain} Piezo1 channel appears as the driver for the stretch-activated current $I_{SAC,NS}$ in a mathematical cardiomyocyte model.
The proposed voltage-modulated Piezo1 formulation reproduces a wide range of experimental results from different groups, including voltage-dependent desensitization and the reversal of desensitization in the presence of positive electrochemical driving forces.
Additionally, after integrating the proposed Piezo1 formulation into the cardiomyocyte model, our studies enable to qualitatively explain some of the experimental observations from~\citet{QuiKoh:2016:cmr}.
First, the loss of mechanical capture in cardiac tissues can -- at least in part -- be attributed to the Piezo1 channel.
Second, it is likely that the voltage-modulation of the Piezo1 channel plays a major role in explaining the  differences in the number of captured mechanical stimuli observed during the electromechanical pacing of cardiac tissues across pacing protocols and frequencies.
However, significant quantitative differences still exist between the number of experimentally observed and numerically simulated captured mechanical stimuli.
In Section~\ref{sec:limitations}, we discuss several reasons for the observed differences and potential avenues to further improve the proposed model.
In addition to improving the proposed model, further mechanisms independent of Piezo1 channels may also play an important role in explaining the experimental observations reported by~\citet{QuiKoh:2016:cmr}.

\subsection{Comparison with related models}

{
Stretch-activated currents at the cellular level have been previously investigated, with a particular emphasis of early studies on the arrhythmogenic effects of these currents. These earlier studies primarily used models based on a single linear stretch-activated current.
In this context, we refer to the discussion in~\cite{BuoLyoDelHeiLum:2024:ees} for a detailed overview on modeling efforts of stretch-activated currents.
}

{
\citet{NieSmi:2007:mms}~studied the slow force response of rat cardiomyocytes under constant stretch.
Unfortunately, at their time the existence of Piezo1 was not yet known.
Enriching their model by a calcium flux due to Piezo1 could improve their model further by potentially explaining the remaining discrepancy in the intracellular calcium concentration between their model and experimental data.
Although our work is similar in nature, the primary focus of~\citet{NieSmi:2007:mms} was to investigate the force response to permanent stretch while we are primarily interested in the interaction between stretch, transmembrane potential, and ionic balances in terms of depolarization inducibility due to stretch. This interaction is particularly important in CPR studies and could be relevant in other heart failure-related mechanisms~(see, e.g.,~\cite{LiaHuaYuaCheLiaZenZheCaoGenZho:2017:scp,ZhaSuLiMaSheWanSheCheJiXieMaXia:2021:pmp}).
}

{
The recent work by~\citet{BuoLyoDelHeiLum:2024:ees} studied stretch-activated ionic currents more systematically with a state-of-the-art human ventricular cardiomyocyte model (ToR-ORd~\cite{TomBuePasZhoMinBriBarSevShrVirVarRod:2019:dcv}).
Their experimental work suggests that the reversal potential for the currents generated by non-selective stretch-activated ion channels is around $\SI{-15}{\milli\volt}$, which differs from the usually assumed $\SI{0}{\milli\volt}$.
However, although their work is complementary to our work in the sense that they present, among other results, calibration procedures for stretch-activated currents, we could not integrate the data from their work directly to enhance our model, since we lack rabbit-specific data for the Piezo1 ion channel.
}

\subsection{Limitations}
\label{sec:limitations}
It has been shown in several experiments that repeated mechanical stimuli can lead to inactivation~(e.g.,~\cite{CosMatSchEarRanPetDubPat:2010:ppa,GotBaeSac:2012:gmc,LewCuiMcDGra:2017:trm,MorSerFleSanLew:2018:vgm}).
Experiments of~\citet{MorSerFleSanLew:2018:vgm} suggest that the inactivation can be reverted with an outward flux of ions in symmetric $K^+$ buffers and positive transmembrane voltages.
However, this setup, which we used as ground truth data to fit the ion channel model's kinetic parameters, has some limitations.
Symmetric $K^+$ buffers are not the cells' physiological environment.
It is hence unclear whether the observed kinetics in~\cite{MorSerFleSanLew:2018:vgm} is directly translatable to more complex ionic environments.
For example, inactivation of mouse Piezo1 channels can be observed at positive transmembrane potentials in more complex ionic environments~\cite[Fig.~2A]{GnaBaeGotSac:2015:isp}.
Hence, some of the discrepancies in the number of captured stimuli simulated with our proposed cardiomyocyte model and observed in the electromechanical pacing protocols may depend on the simplifications made (e.g., the calibration of the kinetic parameters) when integrating the Piezo1 ion channel into the Mahajan-Shiferaw cell model.
Currently, sufficient quantitative experimental data is not available to develop a more precise description of electrochemical driving force of our Piezo1 ion channel model. Hence, we decided to develop this first model with a simplistic description of the before-mentioned electrochemical driving forces (cf. eq.~\ref{eq:IPz1}).

A second problem we encountered in developing the proposed model is the lack of publicly available experimental data specific to the rabbit Piezo1 electrophysiology kinetics.
The parameters for our mathematical model have been estimated using mouse neuroblastoma (N2a) derived Piezo1 kinetics by~\cite{MorSerFleSanLew:2018:vgm}.
Our model is able to describe qualitatively well all the experimental observations regarding the wild-type Piezo1 channel.
However, since there are physiological differences between mouse and rabbit, as well as between neurons and cardiomyocytes, there is the possibility that the rabbit Piezo1 ion channel kinetics differs sufficiently to cause significant deviations in the modeled response.
This hypothesis is supported by recent work on comparing Piezo1 ion channels in bone marrow with those in cardiac tissue~\cite{SimKleEmiChaGreGruLotHilRogRavKohSchPey:2024:psc}.
Furthermore, as already discussed in sec~\ref{sec:molecular-candidate}, experiments suggest that Piezo1 ion channels colocalize with TRP channels~\cite{JiaYinWuZhaWanCheZhoXia:2021:mpc,YuGonKesGuoWuLiCheZhoIisKaiGraCoxFenMar:2022:pcm}.
{It has also been recently shown that Piezo1 can directly interact with TREK1~\cite{LewCroGra:2024:pic}, TRPV~\cite{AllBonTawKamHorMadWad:2025:vep} and TRPM~\cite{GuoCheYuSchChaHilPeyFenCoxMar:2024:fcp}.}
The potential interaction between these ion channels will likely modify their kinetics and hence the generated ionic fluxes.
It has also been demonstrated in the paper by~\citet{MorSerFleSanLew:2018:vgm} that there are observable differences in Piezo1 ion channel kinetics across the different kingdoms. 
Additional experiments may help overcoming these limitations. 
In particular, studies using cardiomyocyte-derived Piezo1 patches and whole-cell experiments 
can help to better understand the role of Piezo1 in the observed loss of capture and other cellular processes.

Another component in our model that needs further investigation is the electrochemical driving force $\Delta\mu$. In our Piezo1 ion channel model, we assume that  $\Delta\mu$ is equivalent to the transmembrane voltage. 
We made this assumption based on experimental observations showing that the reversal potential of the Piezo1 channel is close to zero~\cite{CosMatSchEarRanPetDubPat:2010:ppa,CosXiaSanSyeGraSpeKimSchMatDubMonPat:2012:ppa,GnaBaeGotSac:2015:isp,Got:2017:tf}.
However, we also construct the transmembrane current as the sum of three independent Ohmic currents with separate driving forces (cf. eq.~\eqref{eq:IPz1}), which are not coupled to the driving force in the Piezo1 channel.
This means that the effect of the driving forces may not match with the reversal potential in the Piezo1 channel being zero.
{However, we have verified in supplementary computational studies (not shown) that the corresponding GHK potential remains close to \SI{0}{\milli\volt} in all adopted protocols.}
The motivation for the separation of the total Piezo1 current into the Ohmic currents is simple.
Since Piezo1 generates enough current to depolarize the membrane, there has to be also significant ionic flux, and hence, since the number of different ions has to be conserved in our model, we had to adjust the concentration balance equations.
However, in the Mahajan-Shiferaw model, the balance equations require separate currents, and hence we decided to additively split the current as described. 
This represents the simplest possible formulation since potentially nonlinear or multiplicative couplings are not included. 
{Furthermore, when studying the effect of passive filling on the new ion channel model (see Fig.~\ref{fig:cell-cycle-pacing}) we noticed that the probability mainly concentrates in the slow inactivating state $I_2$ and does not reset as expected from the observations in~\cite{MorSerFleSanLew:2018:vgm}.
This behavior can be reproduced by pacing only mechanically the ion channel at around \SI{10}{\milli\volt} (not shown).
Investigating this behavior further, we noticed that a lower reversal potential of at least $\SI{-30}{\milli\volt}$ is necessary for the ion channel to significantly reset in this setup.
This also explains the high scaling factor in Fig.~\ref{fig:quinn-experiment-loss}, which suggests that an excessive number of Piezo1 ion channels is necessary in the experimental setup.
This discrepancy might be due to several factors.
First, recent experiments studying non-specific stretch activated currents~\cite{BuoLyoDelHeiLum:2024:ees} suggest that the reversal potential could be around \SI{-15}{\milli\volt} (in contrast to the commonly assumed \SI{0}{\milli\volt} used in our model).
Second, although the experiments by~\citet{MorSerFleSanLew:2018:vgm} already provide significant data that we used to fit our model, no substantial information is available regarding the behavior of the Piezo1 ion channel in more complex scenarios as, for example, how the channel will reset at $\approx$\SI{0}{\milli\volt} and how the interaction between the electrochemical gradient $\Delta\mu$ and pressure (p) is for pressures different from \SI{70}{\mmHg}.
Third, we also neglected other stretch-activated currents to keep the parameter space of computational studies tractable.
We have made this decision because we already have insufficient data to correctly calibrate the current for the Piezo1 ion channel in isolation.
Future work should revisit the reversal potential and kinetics when more electromechanical experimental data on cardiomyocytes becomes available.
}

In investigating the electromechanical pacing of cardiac tissue by~\citet{QuiKoh:2016:cmr}, we have introduced the simplification of reproducing these experimental results at the material point level.
In doing so, we ignored any potential influence of the heart's natural electrical pacing (e.g., pacing from the SA node). 
This confounds the comparison of the loss of capture in our simulations and the experiments, as the intracellular ionic concentrations of the cardiomyocyte model evolve differently than in the experiment where the natural electrical pacing is present.
Second, on a structural level, the piston in the experiments generates inhomogeneous pressure gradients in the Langendorff perfused heart.
Since in our setup we study single material points, this effect is a priori not captured.
We do not expect that these simplifications will significantly affect the presented results from a qualitative point of view.
More significant in these simulations can be the effect of the contractile behavior of the cardiomyocytes, which has been neglected in the simulated response focusing on the material point. 
However, these would significantly depend on the structural response, which in turn would depend on boundary conditions associated, e.g., with the unknown pressure in the intraventricular balloon. 
The related changes in tissue pressure and stretch would impact the response of the Piezo1 ion channel.

{
Finally, related to the last point, we have not performed full organ simulations of the heart, but used a simplified setup to study the interaction between passive stretch and cell electrophysiology.
Published computational studies with organ-level electromechanical models make the assumption that only fiber stretch leads to open stretch-activated ion channels, i.e., implicitly assuming that lateral cardiomyocyte stretch can be neglected.
However, we could not find experimental evidence supporting or against this hypothesis. Therefore, before applying the proposed eletromechanical model at the organ level, a better understanding of the mechanism for the translation of macroscopic stretch measures to microscopic ion channels states is needed.
We plan to address this question in subsequent work, after the remaining challenges with respect to integrating, calibrating, and validating the Piezo1 model at the cellular level are resolved.
}

\subsection{Conclusions}

We presented a mathematical model and numerical simulations that integrate mechano-electrical and mechano-chemical feedback at the cellular level. 
This is a first step toward a mechanistic understanding of emergency pacing as well as physiological pathways governing growth and remodeling processes at the cellular scale.

In particular, to understand emergency pacing procedures in detail, more studies have to be conducted at the cellular level in tandem with modelers, as these cellular models bridge the microscopic protein scale with the macroscopic tissue and organ scales.
Subjects receiving an emergency pacing procedure may have underlying disease conditions leading to the cardiac arrest.
Cardiac arrest can also be triggered chemically (e.g., due to substance abuse, substance intolerance or accidental exposure to cardiotoxic substances), electrically, or mechanically via blunt force, as for example in commotio cordis, leading potentially to the application of an emergency pacing procedure.
Future studies on the topic of emergency pacing protocols should take these different circumstances of cardiac arrest into account when deriving hypotheses, constructing enhanced models, and setting up simulation experiments.
We hope that the presented framework will provide a solid first step to develop subsequent studies in the broader context of electromechanical pacing.
By sharing our baseline model in the form of CellML files, we would like to encourage other groups to use our framework to derive more refined integrations into cardiomyocytes, and potentially other cell models where Piezo1 ion channels play an important functional role.

\section*{Acknowledgement}

Financial funding by the German Research Foundation (Deutsche Forschungsgemeinschaft, DFG), project ID 544827709, is highly appreciated by the authors D. Balzani and D. Ogiermann.

\section*{Data availability statement}

The CellML files related to this work have been uploaded to the CellML repository.

\section*{Competing interests}

The authors declare that they have no competing interests.

\bibliographystyle{plainnat}
\bibliography{references}

\begin{thebibliography}{90}
\providecommand{\natexlab}[1]{#1}
\providecommand{\url}[1]{\texttt{#1}}
\expandafter\ifx\csname urlstyle\endcsname\relax
  \providecommand{\doi}[1]{doi: #1}\else
  \providecommand{\doi}{doi: \begingroup \urlstyle{rm}\Url}\fi

\bibitem[Allerkamp et~al.(2025)Allerkamp, Bondarenko, Tawfik, {Kamali-Simsek},
  Horvat~Mercnik, {Madreiter-Sokolowski}, and
  Wadsack]{AllBonTawKamHorMadWad:2025:vep}
Hanna~H. Allerkamp, Alexander~I. Bondarenko, Ines Tawfik, Nil{\"u}fer
  {Kamali-Simsek}, Monika Horvat~Mercnik, Corina~T. {Madreiter-Sokolowski}, and
  Christian Wadsack.
\newblock In vitro examination of {{Piezo1-TRPV4}} dynamics: Implications for
  placental endothelial function in normal and preeclamptic pregnancies.
\newblock \emph{American Journal of Physiology-Cell Physiology}, 328\penalty0
  (1):\penalty0 C227--C244, January 2025.
\newblock ISSN 0363-6143, 1522-1563.
\newblock \doi{10.1152/ajpcell.00794.2024}.

\bibitem[Aronis et~al.(2019)Aronis, Ali, and Trayanova]{AroAliTra:2019:rpa}
Konstantinos~N. Aronis, Rheeda Ali, and Natalia~A. Trayanova.
\newblock The role of personalized atrial modeling in understanding atrial
  fibrillation mechanisms and improving treatment.
\newblock \emph{International Journal of Cardiology}, 287:\penalty0 139--147,
  July 2019.
\newblock ISSN 01675273.
\newblock \doi{10.1016/j.ijcard.2019.01.096}.

\bibitem[Augustin et~al.(2021)Augustin, Gsell, Karabelas, Willemen, Prinzen,
  Lumens, Vigmond, and Plank]{AugGseKarWilPriLumVigPla:2021:cep}
Christoph~M. Augustin, Matthias A.~F. Gsell, Elias Karabelas, Erik Willemen,
  Frits~W. Prinzen, Joost Lumens, Edward~J. Vigmond, and Gernot Plank.
\newblock A computationally efficient physiologically comprehensive
  {{3D}}--{{0D}} closed-loop model of the heart and circulation.
\newblock \emph{Computer Methods in Applied Mechanics and Engineering},
  386:\penalty0 114092, December 2021.
\newblock ISSN 0045-7825.
\newblock \doi{10.1016/j.cma.2021.114092}.

\bibitem[Backs(2022)]{Bac:2022:plm}
Johannes Backs.
\newblock Piezo1 links mechanosensation to cardiac growth.
\newblock \emph{Nature Cardiovascular Research}, 1\penalty0 (6):\penalty0
  533--534, June 2022.
\newblock ISSN 2731-0590.
\newblock \doi{10.1038/s44161-022-00084-y}.

\bibitem[Bae et~al.(2013{\natexlab{a}})Bae, Gnanasambandam, Nicolai, Sachs, and
  Gottlieb]{BaeGnaNicSacGot:2013:xcm}
Chilman Bae, Radhakrishnan Gnanasambandam, Chris Nicolai, Frederick Sachs, and
  Philip~A. Gottlieb.
\newblock Xerocytosis is caused by mutations that alter the kinetics of the
  mechanosensitive channel {{PIEZO1}}.
\newblock \emph{Proceedings of the National Academy of Sciences}, 110\penalty0
  (12), March 2013{\natexlab{a}}.
\newblock ISSN 0027-8424, 1091-6490.
\newblock \doi{10.1073/pnas.1219777110}.

\bibitem[Bae et~al.(2013{\natexlab{b}})Bae, Gottlieb, and
  Sachs]{BaeGotSac:2013:hpr}
Chilman Bae, Philip~A. Gottlieb, and Frederick Sachs.
\newblock Human {{PIEZO1}}: {{Removing Inactivation}}.
\newblock \emph{Biophysical Journal}, 105\penalty0 (4):\penalty0 880--886,
  August 2013{\natexlab{b}}.
\newblock ISSN 00063495.
\newblock \doi{10.1016/j.bpj.2013.07.019}.

\bibitem[Bartoli et~al.(2022)Bartoli, Evans, Blythe, Stewart,
  {Chuntharpursat-Bon}, Debant, Musialowski, Lichtenstein, Parsonage, Futers,
  Turner, and Beech]{BarEvaBlySteChuDebMusLicParFutTurBee:2022:gpg}
Fiona Bartoli, Elizabeth~L. Evans, Nicola~M. Blythe, Leander Stewart, Eulashini
  {Chuntharpursat-Bon}, Marjolaine Debant, Katie~E. Musialowski, Laeticia
  Lichtenstein, Gregory Parsonage, T.~Simon Futers, Neil~A. Turner, and
  David~J. Beech.
\newblock Global {{PIEZO1 Gain-of-Function Mutation Causes Cardiac
  Hypertrophy}} and {{Fibrosis}} in {{Mice}}.
\newblock \emph{Cells}, 11\penalty0 (7):\penalty0 1199, April 2022.
\newblock ISSN 2073-4409.
\newblock \doi{10.3390/cells11071199}.

\bibitem[Buonocunto et~al.(2024)Buonocunto, Lyon, Delhaas, Heijman, and
  Lumens]{BuoLyoDelHeiLum:2024:ees}
Melania Buonocunto, Aurore Lyon, Tammo Delhaas, Jordi Heijman, and Joost
  Lumens.
\newblock Electrophysiological effects of stretch-activated ion channels: A
  systematic computational characterization.
\newblock \emph{The Journal of Physiology}, 602\penalty0 (18):\penalty0
  4585--4604, September 2024.
\newblock ISSN 0022-3751, 1469-7793.
\newblock \doi{10.1113/JP284439}.

\bibitem[Cahalan et~al.(2015)Cahalan, Lukacs, Ranade, Chien, Bandell, and
  Patapoutian]{CahLukRanChiBanPat:2015:plm}
Stuart~M Cahalan, Viktor Lukacs, Sanjeev~S Ranade, Shu Chien, Michael Bandell,
  and Ardem Patapoutian.
\newblock Piezo1 links mechanical forces to red blood cell volume.
\newblock \emph{eLife}, 4:\penalty0 e07370, May 2015.
\newblock ISSN 2050-084X.
\newblock \doi{10.7554/eLife.07370}.

\bibitem[Camps et~al.(2020)Camps, Lawson, Drovandi, Minchole, Wang, Grau,
  Burrage, and Rodriguez]{CamLawDroMinWanGraBurRod:2020:Iva}
Julia Camps, Brodie Lawson, Christopher Drovandi, Ana Minchole, Zhinuo~Jenny
  Wang, Vicente Grau, Kevin Burrage, and Blanca Rodriguez.
\newblock Inference of ventricular activation properties from non-invasive
  electrocardiography.
\newblock \emph{arXiv:2010.15214 [eess, q-bio]}, October 2020.

\bibitem[{Cardone-Noott} et~al.(2016){Cardone-Noott}, {Bueno-Orovio},
  Minchol{\'e}, Zemzemi, and Rodriguez]{CarBueMinZemRod:2016:Hva}
Louie {Cardone-Noott}, Alfonso {Bueno-Orovio}, Ana Minchol{\'e}, Nejib Zemzemi,
  and Blanca Rodriguez.
\newblock Human ventricular activation sequence and the simulation of the
  electrocardiographic {{QRS}} complex and its variability in healthy and
  intraventricular block conditions.
\newblock \emph{EP Europace}, 18\penalty0 (suppl\_4):\penalty0 iv4--iv15,
  December 2016.
\newblock \doi{10.1093/europace/euw346}.

\bibitem[Celaya et~al.(2014)Celaya, Aguirrezabala, and
  Chatzipantelidis]{CelAguCha:2014:iab}
E.~Alberdi Celaya, J.~J.~Anza Aguirrezabala, and P.~Chatzipantelidis.
\newblock Implementation of an {{Adaptive BDF2 Formula}} and {{Comparison}}
  with the {{MATLAB Ode15s}}.
\newblock \emph{Procedia Computer Science}, 29:\penalty0 1014--1026, 2014.
\newblock ISSN 18770509.
\newblock \doi{10.1016/j.procs.2014.05.091}.

\bibitem[Chong et~al.(2021)Chong, De~Vecchis, Hyman, Povstyan, Ludlow, Shi,
  Beech, and Kalli]{ChoDeHymPovLudShiBeeKal:2021:mfp}
Jiehan Chong, Dario De~Vecchis, Adam~J. Hyman, Oleksandr~V. Povstyan,
  Melanie~J. Ludlow, Jian Shi, David~J. Beech, and Antreas~C. Kalli.
\newblock Modeling of full-length {{Piezo1}} suggests importance of the
  proximal {{N-terminus}} for dome structure.
\newblock \emph{Biophysical Journal}, 120\penalty0 (8):\penalty0 1343--1356,
  April 2021.
\newblock ISSN 00063495.
\newblock \doi{10.1016/j.bpj.2021.02.003}.

\bibitem[Clerx et~al.(2020)Clerx, Cooling, Cooper, Garny, Moyle, Nickerson,
  Nielsen, and Sorby]{CleCooCooGarMoyNicNieSor:2020:c}
Michael Clerx, Michael~T. Cooling, Jonathan Cooper, Alan Garny, Keri Moyle,
  David~P. Nickerson, Poul M.~F. Nielsen, and Hugh Sorby.
\newblock {{CellML}} 2.0.
\newblock \emph{Journal of Integrative Bioinformatics}, 17\penalty0
  (2-3):\penalty0 20200021, August 2020.
\newblock ISSN 1613-4516.
\newblock \doi{10.1515/jib-2020-0021}.

\bibitem[Colli~Franzone et~al.(2017)Colli~Franzone, Pavarino, and
  Scacchi]{ColPavSca:2017:emf}
P.~Colli~Franzone, L.~F. Pavarino, and S.~Scacchi.
\newblock Effects of mechanical feedback on the stability of cardiac scroll
  waves: {{A}} bidomain electro-mechanical simulation study.
\newblock \emph{Chaos: An Interdisciplinary Journal of Nonlinear Science},
  27\penalty0 (9):\penalty0 093905, September 2017.
\newblock ISSN 1054-1500.
\newblock \doi{10.1063/1.4999465}.

\bibitem[Coste et~al.(2010)Coste, Mathur, Schmidt, Earley, Ranade, Petrus,
  Dubin, and Patapoutian]{CosMatSchEarRanPetDubPat:2010:ppa}
Bertrand Coste, Jayanti Mathur, Manuela Schmidt, Taryn~J. Earley, Sanjeev
  Ranade, Matt~J. Petrus, Adrienne~E. Dubin, and Ardem Patapoutian.
\newblock Piezo1 and {{Piezo2 Are Essential Components}} of {{Distinct
  Mechanically Activated Cation Channels}}.
\newblock \emph{Science}, 330\penalty0 (6000):\penalty0 55--60, October 2010.
\newblock ISSN 0036-8075, 1095-9203.
\newblock \doi{10.1126/science.1193270}.

\bibitem[Coste et~al.(2012)Coste, Xiao, Santos, Syeda, Grandl, Spencer, Kim,
  Schmidt, Mathur, Dubin, Montal, and
  Patapoutian]{CosXiaSanSyeGraSpeKimSchMatDubMonPat:2012:ppa}
Bertrand Coste, Bailong Xiao, Jose~S. Santos, Ruhma Syeda, J{\"o}rg Grandl,
  Kathryn~S. Spencer, Sung~Eun Kim, Manuela Schmidt, Jayanti Mathur,
  Adrienne~E. Dubin, Mauricio Montal, and Ardem Patapoutian.
\newblock Piezo proteins are pore-forming subunits of mechanically activated
  channels.
\newblock \emph{Nature}, 483\penalty0 (7388):\penalty0 176--181, March 2012.
\newblock ISSN 0028-0836, 1476-4687.
\newblock \doi{10.1038/nature10812}.

\bibitem[De~Vecchis et~al.(2021)De~Vecchis, Beech, and
  Kalli]{DeBeeKal:2021:mds}
Dario De~Vecchis, David~J. Beech, and Antreas~C. Kalli.
\newblock Molecular dynamics simulations of {{Piezo1}} channel opening by
  increases in membrane tension.
\newblock \emph{Biophysical Journal}, 120\penalty0 (8):\penalty0 1510--1521,
  April 2021.
\newblock ISSN 00063495.
\newblock \doi{10.1016/j.bpj.2021.02.006}.

\bibitem[Deng et~al.(2016)Deng, Arevalo, Prakosa, Callans, and
  Trayanova]{DenArePraCalTra:2016:fsa}
Dongdong Deng, Hermenegild~J. Arevalo, Adityo Prakosa, David~J. Callans, and
  Natalia~A. Trayanova.
\newblock A feasibility study of arrhythmia risk prediction in patients with
  myocardial infarction and preserved ejection fraction.
\newblock \emph{EP Europace}, 18\penalty0 (suppl\_4):\penalty0 iv60--iv66,
  December 2016.
\newblock ISSN 1099-5129, 1532-2092.
\newblock \doi{10.1093/europace/euw351}.

\bibitem[Eyring(1935{\natexlab{a}})]{Eyr:1935:aca}
{\relax Henry}.~Eyring.
\newblock The {{Activated Complex}} and the {{Absolute Rate}} of {{Chemical
  Reactions}}.
\newblock \emph{Chemical Reviews}, 17\penalty0 (1):\penalty0 65--77, August
  1935{\natexlab{a}}.
\newblock ISSN 0009-2665, 1520-6890.
\newblock \doi{10.1021/cr60056a006}.

\bibitem[Eyring(1935{\natexlab{b}})]{Eyr:1935:acc}
Henry Eyring.
\newblock The {{Activated Complex}} in {{Chemical Reactions}}.
\newblock \emph{The Journal of Chemical Physics}, 3\penalty0 (2):\penalty0
  107--115, February 1935{\natexlab{b}}.
\newblock ISSN 0021-9606, 1089-7690.
\newblock \doi{10.1063/1.1749604}.

\bibitem[Fedele et~al.(2023)Fedele, Piersanti, Regazzoni, Salvador, Africa,
  Bucelli, Zingaro, Dede', and
  Quarteroni]{FedPieRegSalAfrBucZinDedQua:2023:cbd}
Marco Fedele, Roberto Piersanti, Francesco Regazzoni, Matteo Salvador,
  Pasquale~Claudio Africa, Michele Bucelli, Alberto Zingaro, Luca Dede', and
  Alfio Quarteroni.
\newblock A comprehensive and biophysically detailed computational model of the
  whole human heart electromechanics.
\newblock \emph{Computer Methods in Applied Mechanics and Engineering},
  410:\penalty0 115983, May 2023.
\newblock ISSN 00457825.
\newblock \doi{10.1016/j.cma.2023.115983}.

\bibitem[Garny and Hunter(2015)]{GarHun:2015:Omi}
Alan Garny and Peter~J. Hunter.
\newblock {{OpenCOR}}: A modular and interoperable approach to computational
  biology.
\newblock \emph{Frontiers in Physiology}, 6:\penalty0 26, February 2015.
\newblock ISSN 1664-042X.
\newblock \doi{10.3389/fphys.2015.00026}.

\bibitem[Gerach and Loewe(2024)]{GerLoe:2024:dem}
Tobias Gerach and Axel Loewe.
\newblock Differential effects of mechano-electric feedback mechanisms on
  whole-heart activation, repolarization, and tension.
\newblock \emph{The Journal of Physiology}, 602\penalty0 (18):\penalty0
  4605--4624, September 2024.
\newblock ISSN 0022-3751, 1469-7793.
\newblock \doi{10.1113/JP285022}.

\bibitem[Gerach et~al.(2021)Gerach, Schuler, Fr{\"o}hlich, Lindner, Kovacheva,
  Moss, W{\"u}lfers, Seemann, Wieners, and
  Loewe]{GerSchFroLinKovMosWulSeeWieLoe:2021:EWD}
Tobias Gerach, Steffen Schuler, Jonathan Fr{\"o}hlich, Laura Lindner, Ekaterina
  Kovacheva, Robin Moss, Eike~Moritz W{\"u}lfers, Gunnar Seemann, Christian
  Wieners, and Axel Loewe.
\newblock Electro-{{Mechanical Whole-Heart Digital Twins}}: {{A Fully Coupled
  Multi-Physics Approach}}.
\newblock \emph{Mathematics}, 9\penalty0 (11):\penalty0 1247, January 2021.
\newblock \doi{10.3390/math9111247}.

\bibitem[Gerbi et~al.(2019)Gerbi, Ded{\`e}, Quarteroni, Gerbi, Ded{\`e}, and
  Quarteroni]{GerDedQuaGerDedQua:2019:mas}
Antonello Gerbi, Luca Ded{\`e}, Alfio Quarteroni, Antonello Gerbi, Luca
  Ded{\`e}, and Alfio Quarteroni.
\newblock A monolithic algorithm for the simulation of cardiac electromechanics
  in the human left ventricle.
\newblock \emph{Mathematics in Engineering}, 1\penalty0 (1):\penalty0 1--37,
  2019.
\newblock ISSN 2640-3501.
\newblock \doi{10.3934/Mine.2018.1.1}.

\bibitem[Gnanasambandam et~al.(2015)Gnanasambandam, Bae, Gottlieb, and
  Sachs]{GnaBaeGotSac:2015:isp}
Radhakrishnan Gnanasambandam, Chilman Bae, Philip~A. Gottlieb, and Frederick
  Sachs.
\newblock Ionic {{Selectivity}} and {{Permeation Properties}} of {{Human PIEZO1
  Channels}}.
\newblock \emph{PLOS ONE}, 10\penalty0 (5):\penalty0 e0125503, May 2015.
\newblock ISSN 1932-6203.
\newblock \doi{10.1371/journal.pone.0125503}.

\bibitem[G{\"o}ktepe et~al.(2014)G{\"o}ktepe, Menzel, and
  Kuhl]{GokMenKuh:2014:ghm}
Serdar G{\"o}ktepe, Andreas Menzel, and Ellen Kuhl.
\newblock The generalized {{Hill}} model: {{A}} kinematic approach towards
  active muscle contraction.
\newblock \emph{Journal of the Mechanics and Physics of Solids}, 72:\penalty0
  20--39, December 2014.
\newblock ISSN 0022-5096.
\newblock \doi{10.1016/J.JMPS.2014.07.015}.

\bibitem[Gottlieb(2017)]{Got:2017:tf}
P.A. Gottlieb.
\newblock A {{Tour}} de {{Force}}.
\newblock In \emph{Current {{Topics}} in {{Membranes}}}, volume~79, pages
  1--36. Elsevier, 2017.
\newblock ISBN 978-0-12-809389-4.
\newblock \doi{10.1016/bs.ctm.2016.11.007}.

\bibitem[Gottlieb et~al.(2012)Gottlieb, Bae, and Sachs]{GotBaeSac:2012:gmc}
Philip~A. Gottlieb, Chilman Bae, and Frederick Sachs.
\newblock Gating the mechanical channel {{Piezo1}}: {{A}} comparison between
  whole-cell and patch recording.
\newblock \emph{Channels}, 6\penalty0 (4):\penalty0 282--289, July 2012.
\newblock ISSN 1933-6950, 1933-6969.
\newblock \doi{10.4161/chan.21064}.

\bibitem[Grandi et~al.(2019)Grandi, Dobrev, and Heijman]{GraDobHei:2019:cmw}
Eleonora Grandi, Dobromir Dobrev, and Jordi Heijman.
\newblock Computational modeling: {{What}} does it tell us about atrial
  fibrillation therapy?
\newblock \emph{International Journal of Cardiology}, 287:\penalty0 155--161,
  July 2019.
\newblock ISSN 01675273.
\newblock \doi{10.1016/j.ijcard.2019.01.077}.

\bibitem[Guo et~al.(2024)Guo, Cheng, Yu, Schiatti, Chan, Hill, Peyronnet,
  Feneley, Cox, and Martinac]{GuoCheYuSchChaHilPeyFenCoxMar:2024:fcp}
Yang Guo, Delfine Cheng, Ze-Yan Yu, Teresa Schiatti, Andrea~Y. Chan, Adam~P.
  Hill, R{\'e}mi Peyronnet, Michael~P. Feneley, Charles~D. Cox, and Boris
  Martinac.
\newblock Functional coupling between {{Piezo1}} and {{TRPM4}} influences the
  electrical activity of {{HL}}-1 atrial myocytes.
\newblock \emph{The Journal of Physiology}, 602\penalty0 (18):\penalty0
  4363--4386, September 2024.
\newblock ISSN 0022-3751, 1469-7793.
\newblock \doi{10.1113/JP284474}.

\bibitem[Hindmarsh et~al.(2005)Hindmarsh, Brown, Grant, Lee, Serban, Shumaker,
  and Woodward]{HinBroGraLeeSerShuWoo:2005:ssn}
Alan~C. Hindmarsh, Peter~N. Brown, Keith~E. Grant, Steven~L. Lee, Radu Serban,
  Dan~E. Shumaker, and Carol~S. Woodward.
\newblock {{SUNDIALS}}: {{Suite}} of nonlinear and differential/algebraic
  equation solvers.
\newblock \emph{ACM Transactions on Mathematical Software}, 31\penalty0
  (3):\penalty0 363--396, September 2005.
\newblock ISSN 0098-3500, 1557-7295.
\newblock \doi{10.1145/1089014.1089020}.

\bibitem[Huber(1964)]{Hub:1964:rel}
Peter~J. Huber.
\newblock Robust {{Estimation}} of a {{Location Parameter}}.
\newblock \emph{The Annals of Mathematical Statistics}, 35\penalty0
  (1):\penalty0 73--101, March 1964.
\newblock ISSN 0003-4851.
\newblock \doi{10.1214/aoms/1177703732}.

\bibitem[Izu et~al.(2020)Izu, Kohl, Boyden, Miura, Banyasz, Chiamvimonvat,
  Trayanova, Bers, and {Chen-Izu}]{IzuKohBoyMiuBanChiTraBerChe:2020:mmc}
Leighton~T. Izu, Peter Kohl, Penelope~A. Boyden, Masahito Miura, Tamas Banyasz,
  Nipavan Chiamvimonvat, Natalia Trayanova, Donald~M. Bers, and Ye~{Chen-Izu}.
\newblock Mechano-electric and mechano-chemo-transduction in cardiomyocytes.
\newblock \emph{The Journal of Physiology}, 598\penalty0 (7):\penalty0
  1285--1305, 2020.
\newblock ISSN 1469-7793.
\newblock \doi{10.1113/JP276494}.

\bibitem[Jiang et~al.(2021)Jiang, Yin, Wu, Zhang, Wang, Cheng, Zhou, and
  Xiao]{JiaYinWuZhaWanCheZhoXia:2021:mpc}
Fan Jiang, Kunlun Yin, Kun Wu, Mingmin Zhang, Shiqiang Wang, Heping Cheng, Zhou
  Zhou, and Bailong Xiao.
\newblock The mechanosensitive {{Piezo1}} channel mediates heart mechano-chemo
  transduction.
\newblock \emph{Nature Communications}, 12\penalty0 (1):\penalty0 869, December
  2021.
\newblock ISSN 2041-1723.
\newblock \doi{10.1038/s41467-021-21178-4}.

\bibitem[Keener and Sneyd(2009)]{KeeSne:2009:MPC}
J~Keener and J~Sneyd, editors.
\newblock \emph{Mathematical {{Physiology I}}: {{Cellular Physiology}}, {{II}}:
  {{Systems Physiology}}}.
\newblock 2009.

\bibitem[Kloth et~al.(2022)Kloth, Mearini, Weinberger, Stenzig, Geertz,
  Starbatty, Lindner, Schumacher, Reichenspurner, Eschenhagen, and
  Hirt]{KloMeaWeiSteGeeStaLinSchReiEscHir:2022:pni}
Benjamin Kloth, Giulia Mearini, Florian Weinberger, Justus Stenzig, Birgit
  Geertz, Jutta Starbatty, Diana Lindner, Udo Schumacher, Hermann
  Reichenspurner, Thomas Eschenhagen, and Marc~N. Hirt.
\newblock Piezo2 is not an indispensable mechanosensor in murine
  cardiomyocytes.
\newblock \emph{Scientific Reports}, 12\penalty0 (1):\penalty0 8193, December
  2022.
\newblock ISSN 2045-2322.
\newblock \doi{10.1038/s41598-022-12085-9}.

\bibitem[Lee et~al.(2022)Lee, Cans{\i}z, and Kaliske]{LeeCanKal:2022:cmm}
Yongjae Lee, Bar{\i}{\c s} Cans{\i}z, and Michael Kaliske.
\newblock Computational modelling of mechano-electric feedback and its
  arrhythmogenic effects in human ventricular models.
\newblock \emph{Computer Methods in Biomechanics and Biomedical Engineering},
  25\penalty0 (15):\penalty0 1767--1783, November 2022.
\newblock ISSN 1025-5842, 1476-8259.
\newblock \doi{10.1080/10255842.2022.2037573}.

\bibitem[Lewis and Grandl(2021)]{LewGra:2021:pic}
Amanda~H Lewis and J{\"o}rg Grandl.
\newblock Piezo1 ion channels inherently function as independent
  mechanotransducers.
\newblock \emph{eLife}, 10:\penalty0 e70988, October 2021.
\newblock ISSN 2050-084X.
\newblock \doi{10.7554/eLife.70988}.

\bibitem[Lewis et~al.(2017)Lewis, Cui, McDonald, and
  Grandl]{LewCuiMcDGra:2017:trm}
Amanda~H. Lewis, Alisa~F. Cui, Malcolm~F. McDonald, and J{\"o}rg Grandl.
\newblock Transduction of {{Repetitive Mechanical Stimuli}} by {{Piezo1}} and
  {{Piezo2 Ion Channels}}.
\newblock \emph{Cell Reports}, 19\penalty0 (12):\penalty0 2572--2585, June
  2017.
\newblock ISSN 22111247.
\newblock \doi{10.1016/j.celrep.2017.05.079}.

\bibitem[Lewis et~al.(2024)Lewis, Cronin, and Grandl]{LewCroGra:2024:pic}
Amanda~H. Lewis, Marie~E. Cronin, and J{\"o}rg Grandl.
\newblock Piezo1 ion channels are capable of conformational signaling.
\newblock \emph{Neuron}, 112\penalty0 (18):\penalty0 3161--3175.e5, September
  2024.
\newblock ISSN 08966273.
\newblock \doi{10.1016/j.neuron.2024.06.024}.

\bibitem[Li et~al.(2014)Li, Hou, Tumova, Muraki, Bruns, Ludlow, Sedo, Hyman,
  McKeown, Young, Yuldasheva, Majeed, Wilson, Rode, Bailey, Kim, Fu, Carter,
  Bilton, Imrie, Ajuh, Dear, Cubbon, Kearney, Prasad, Evans, Ainscough, and
  Beech]{LiHouTumMurBruLudSedHymMcKYouYulMajWilRodBaiKimFuCarBilImrAjuDeaCubKeaPraEvaAinBee:2014:piv}
Jing Li, Bing Hou, Sarka Tumova, Katsuhiko Muraki, Alexander Bruns, Melanie~J.
  Ludlow, Alicia Sedo, Adam~J. Hyman, Lynn McKeown, Richard~S. Young, Nadira~Y.
  Yuldasheva, Yasser Majeed, Lesley~A. Wilson, Baptiste Rode, Marc~A. Bailey,
  Hyejeong~R. Kim, Zhaojun Fu, Deborah A.~L. Carter, Jan Bilton, Helen Imrie,
  Paul Ajuh, T.~Neil Dear, Richard~M. Cubbon, Mark~T. Kearney, K.~Raj Prasad,
  Paul~C. Evans, Justin F.~X. Ainscough, and David~J. Beech.
\newblock Piezo1 integration of vascular architecture with physiological force.
\newblock \emph{Nature}, 515\penalty0 (7526):\penalty0 279--282, November 2014.
\newblock ISSN 0028-0836, 1476-4687.
\newblock \doi{10.1038/nature13701}.

\bibitem[Liang et~al.(2017)Liang, Huang, Yuan, Chen, Liang, Zeng, Zheng, Cao,
  Geng, and Zhou]{LiaHuaYuaCheLiaZenZheCaoGenZho:2017:scp}
Jianlin Liang, Boshui Huang, Guiyi Yuan, Ying Chen, Fasheng Liang, Huayuan
  Zeng, Shaoxin Zheng, Liang Cao, Dengfeng Geng, and Shuxian Zhou.
\newblock Stretch-activated channel {{Piezo1}} is up-regulated in failure heart
  and cardiomyocyte stimulated by {{AngII}}.
\newblock \emph{American Journal of Translational Research}, 9\penalty0
  (6):\penalty0 2945--2955, 2017.
\newblock ISSN 1943-8141.

\bibitem[Lim(2022)]{Lim:2022:psp}
Gregory~B. Lim.
\newblock Piezo1 senses pressure overload and initiates cardiac hypertrophy.
\newblock \emph{Nature Reviews Cardiology}, 19\penalty0 (8):\penalty0 503--503,
  August 2022.
\newblock ISSN 1759-5002, 1759-5010.
\newblock \doi{10.1038/s41569-022-00746-1}.

\bibitem[Liu and Nocedal(1989)]{LiuNoc:1989:lmb}
Dong~C. Liu and Jorge Nocedal.
\newblock On the limited memory {{BFGS}} method for large scale optimization.
\newblock \emph{Mathematical Programming}, 45\penalty0 (1-3):\penalty0
  503--528, August 1989.
\newblock ISSN 0025-5610, 1436-4646.
\newblock \doi{10.1007/BF01589116}.

\bibitem[Loman et~al.(2022)Loman, Ma, Ilin, Gowda, Korsbo, Yewale, Rackauckas,
  and Isaacson]{LomMaIliGowKorYewRacIsa:2022:cfb}
Torkel~E. Loman, Yingbo Ma, Vasily Ilin, Shashi Gowda, Niklas Korsbo, Nikhil
  Yewale, Chris Rackauckas, and Samuel~A. Isaacson.
\newblock Catalyst: {{Fast Biochemical Modeling}} with {{Julia}}.
\newblock Preprint, Systems Biology, August 2022.

\bibitem[Ma et~al.(2021)Ma, Gowda, Anantharaman, Laughman, Shah, and
  Rackauckas]{MaGowAnaLauShaRac:2021:mcg}
Yingbo Ma, Shashi Gowda, Ranjan Anantharaman, Chris Laughman, Viral Shah, and
  Chris Rackauckas.
\newblock {{ModelingToolkit}}: {{A Composable Graph Transformation System For
  Equation-Based Modeling}}, March 2021.

\bibitem[Mahajan et~al.(2008)Mahajan, Shiferaw, Sato, Baher, Olcese, Xie, Yang,
  Chen, Restrepo, Karma, Garfinkel, Qu, and
  Weiss]{MahShiSatBahOlcXieYanCheResKarGarQuWei:2008:rva}
Aman Mahajan, Yohannes Shiferaw, Daisuke Sato, Ali Baher, Riccardo Olcese,
  Lai-Hua Xie, Ming-Jim Yang, Peng-Sheng Chen, Juan~G. Restrepo, Alain Karma,
  Alan Garfinkel, Zhilin Qu, and James~N. Weiss.
\newblock A {{Rabbit Ventricular Action Potential Model Replicating Cardiac
  Dynamics}} at {{Rapid Heart Rates}}.
\newblock \emph{Biophysical Journal}, 94\penalty0 (2):\penalty0 392--410,
  January 2008.
\newblock ISSN 0006-3495.
\newblock \doi{10.1529/BIOPHYSJ.106.98160}.

\bibitem[Maillet et~al.(2013)Maillet, {van Berlo}, and
  Molkentin]{MaivanMol:2013:mbp}
Marjorie Maillet, Jop~H. {van Berlo}, and Jeffery~D. Molkentin.
\newblock Molecular basis of physiological heart growth: Fundamental concepts
  and new players.
\newblock \emph{Nature Reviews Molecular Cell Biology}, 14\penalty0
  (1):\penalty0 38--48, January 2013.
\newblock ISSN 1471-0072, 1471-0080.
\newblock \doi{10.1038/nrm3495}.

\bibitem[McMillin et~al.(2014)McMillin, Beck, Chong, Shively, Buckingham,
  Gildersleeve, Aracena, Aylsworth, Bitoun, Carey, Clericuzio, Crow, Curry,
  Devriendt, Everman, Fryer, Gibson, Giovannucci~Uzielli, Graham, Hall, Hecht,
  Heidenreich, Hurst, Irani, Krapels, Leroy, Mowat, Plant, Robertson, Schorry,
  Scott, Seaver, Sherr, Splitt, Stewart, Stumpel, Temel, Weaver, Whiteford,
  Williams, Tabor, Smith, Shendure, Nickerson, and
  Bamshad]{McMBecChoShiBucGilAraAylBitCarCleCroCurDevEveFryGibGioGraHalHecHeiHurIraKraLerMowPlaRobSchScoSeaSheSplSteStuTemWeaWhiWilTabSmiSheNicBam:2014:mpc}
Margaret~J. McMillin, Anita~E. Beck, Jessica~X. Chong, Kathryn~M. Shively,
  Kati~J. Buckingham, Heidi~I.S. Gildersleeve, Mariana~I. Aracena, Arthur~S.
  Aylsworth, Pierre Bitoun, John~C. Carey, Carol~L. Clericuzio, Yanick~J. Crow,
  Cynthia~J. Curry, Koenraad Devriendt, David~B. Everman, Alan Fryer, Kate
  Gibson, Maria~Luisa Giovannucci~Uzielli, John~M. Graham, Judith~G. Hall,
  Jacqueline~T. Hecht, Randall~A. Heidenreich, Jane~A. Hurst, Sarosh Irani,
  Ingrid~P.C. Krapels, Jules~G. Leroy, David Mowat, Gordon~T. Plant, Stephen~P.
  Robertson, Elizabeth~K. Schorry, Richard~H. Scott, Laurie~H. Seaver, Elliott
  Sherr, Miranda Splitt, Helen Stewart, Constance Stumpel, Sehime~G. Temel,
  David~D. Weaver, Margo Whiteford, Marc~S. Williams, Holly~K. Tabor, Joshua~D.
  Smith, Jay Shendure, Deborah~A. Nickerson, and Michael~J. Bamshad.
\newblock Mutations in {{PIEZO2 Cause Gordon Syndrome}}, {{Marden-Walker
  Syndrome}}, and {{Distal Arthrogryposis Type}} 5.
\newblock \emph{The American Journal of Human Genetics}, 94\penalty0
  (5):\penalty0 734--744, May 2014.
\newblock ISSN 00029297.
\newblock \doi{10.1016/j.ajhg.2014.03.015}.

\bibitem[Moroni et~al.(2018)Moroni, {Servin-Vences}, Fleischer,
  {S{\'a}nchez-Carranza}, and Lewin]{MorSerFleSanLew:2018:vgm}
Mirko Moroni, M.~Rocio {Servin-Vences}, Raluca Fleischer, Oscar
  {S{\'a}nchez-Carranza}, and Gary~R. Lewin.
\newblock Voltage gating of mechanosensitive {{PIEZO}} channels.
\newblock \emph{Nature Communications}, 9\penalty0 (1):\penalty0 1096, December
  2018.
\newblock ISSN 2041-1723.
\newblock \doi{10.1038/s41467-018-03502-7}.

\bibitem[Moss et~al.(2022)Moss, W{\"u}lfers, Lewetag, Hornyik, {Perez-Feliz},
  Strohbach, Menza, Krafft, Odening, and
  Seemann]{MosWulLewHorPerStrMenKraOdeSee:2022:cmr}
Robin Moss, Eike~M. W{\"u}lfers, Raphaela Lewetag, Tibor Hornyik, Stefanie
  {Perez-Feliz}, Tim Strohbach, Marius Menza, Axel Krafft, Katja~E. Odening,
  and Gunnar Seemann.
\newblock A computational model of rabbit geometry and {{ECG}}: {{Optimizing}}
  ventricular activation sequence and {{APD}} distribution.
\newblock \emph{PLOS ONE}, 17\penalty0 (6):\penalty0 e0270559, June 2022.
\newblock ISSN 1932-6203.
\newblock \doi{10.1371/journal.pone.0270559}.

\bibitem[Mulhall et~al.(2023)Mulhall, Gharpure, Lee, Dubin, Aaron, Marshall,
  Spencer, Reiche, Henderson, Chew, and
  Patapoutian]{MulGhaLeeDubAarMarSpeReiHenChePat:2023:doc}
Eric~M. Mulhall, Anant Gharpure, Rachel~M. Lee, Adrienne~E. Dubin, Jesse~S.
  Aaron, Kara~L. Marshall, Kathryn~R. Spencer, Michael~A. Reiche, Scott~C.
  Henderson, Teng-Leong Chew, and Ardem Patapoutian.
\newblock Direct observation of the conformational states of {{PIEZO1}}.
\newblock \emph{Nature}, 620\penalty0 (7976):\penalty0 1117--1125, August 2023.
\newblock ISSN 0028-0836, 1476-4687.
\newblock \doi{10.1038/s41586-023-06427-4}.

\bibitem[Niederer and Smith(2007)]{NieSmi:2007:mms}
Steven~A. Niederer and Nicolas~P. Smith.
\newblock A {{Mathematical Model}} of the {{Slow Force Response}} to
  {{Stretch}} in {{Rat Ventricular Myocytes}}.
\newblock \emph{Biophysical Journal}, 92\penalty0 (11):\penalty0 4030--4044,
  June 2007.
\newblock ISSN 0006-3495.
\newblock \doi{10.1529/biophysj.106.095463}.

\bibitem[Nikolaev et~al.(2019)Nikolaev, Cox, Ridone, Rohde, {Cordero-Morales},
  V{\'a}squez, Laver, and Martinac]{NikCoxRidRohCorVasLavMar:2019:mti}
Y.~A. Nikolaev, C.~D. Cox, P.~Ridone, P.~R. Rohde, J.~F. {Cordero-Morales},
  V.~V{\'a}squez, D.~R. Laver, and B.~Martinac.
\newblock Mammalian {{TRP}} ion channels are insensitive to membrane stretch.
\newblock \emph{Journal of Cell Science}, page jcs.238360, January 2019.
\newblock ISSN 1477-9137, 0021-9533.
\newblock \doi{10.1242/jcs.238360}.

\bibitem[Ogiermann et~al.(2021)Ogiermann, Balzani, and
  Perotti]{OgiBalPer:2021:ema}
Dennis Ogiermann, Daniel Balzani, and Luigi~E. Perotti.
\newblock The {{Effect}} of {{Modeling Assumptions}} on the {{ECG}} in
  {{Monodomain}} and {{Bidomain Simulations}}.
\newblock In Daniel~B. Ennis, Luigi~E. Perotti, and Vicky~Y. Wang, editors,
  \emph{Functional {{Imaging}} and {{Modeling}} of the {{Heart}}}, volume
  12738, pages 503--514. Springer International Publishing, Cham, 2021.
\newblock ISBN 978-3-030-78709-7 978-3-030-78710-3.
\newblock \doi{$10.1007/978-3-030-78710-3_48$}.

\bibitem[Ogiermann et~al.(2023)Ogiermann, Balzani, and
  Perotti]{OgiBalPer:2023:egh}
Dennis Ogiermann, Daniel Balzani, and Luigi~E. Perotti.
\newblock An extended generalized hill model for cardiac tissue: {{Comparison}}
  with different approaches based on experimental data.
\newblock In \emph{Functional {{Imaging}} and {{Modeling}} of the {{Heart}}},
  pages 555--564, Cham, June 2023. Springer Nature Switzerland.
\newblock ISBN 978-3-031-35302-4.

\bibitem[Olasveengen et~al.(2020)Olasveengen, Mancini, Perkins, Avis, Brooks,
  Castr{\'e}n, Chung, Considine, Couper, Escalante, Hatanaka, Hung, Kudenchuk,
  Lim, Nishiyama, Ristagno, Semeraro, Smith, Smyth, Vaillancourt, Nolan,
  Hazinski, Morley, Svavarsd{\'o}ttir, Raffay, Kuzovlev, Grasner, Dee, Smith,
  and
  Rajendran]{OlaManPerAviBroCasChuConCouEscHatHunKudLimNisRisSemSmiSmyVaiNolHazMorSvaRafKuzGraDeeSmiRaj:2020:abl}
Theresa~M. Olasveengen, Mary~E. Mancini, Gavin~D. Perkins, Suzanne Avis, Steven
  Brooks, Maaret Castr{\'e}n, Sung~Phil Chung, Julie Considine, Keith Couper,
  Raffo Escalante, Tetsuo Hatanaka, Kevin~K.C. Hung, Peter Kudenchuk, Swee~Han
  Lim, Chika Nishiyama, Giuseppe Ristagno, Federico Semeraro, Christopher~M.
  Smith, Michael~A. Smyth, Christian Vaillancourt, Jerry~P. Nolan, Mary~Fran
  Hazinski, Peter~T. Morley, Hildigunnur Svavarsd{\'o}ttir, Violetta Raffay,
  Artem Kuzovlev, Jan-Thorsten Grasner, Ryan Dee, Michael Smith, and Kausala
  Rajendran.
\newblock Adult {{Basic Life Support}}: 2020 {{International Consensus}} on
  {{Cardiopulmonary Resuscitation}} and {{Emergency Cardiovascular Care Science
  With Treatment Recommendations}}.
\newblock \emph{Circulation}, 142\penalty0 (16\_suppl\_1), October 2020.
\newblock ISSN 0009-7322, 1524-4539.
\newblock \doi{10.1161/CIR.0000000000000892}.

\bibitem[Passini et~al.(2017)Passini, Britton, Lu, Rohrbacher, Hermans,
  Gallacher, Greig, {Bueno-Orovio}, and
  Rodriguez]{PasBriLuRohHerGalGreBueRod:2017:hsd}
Elisa Passini, Oliver~J. Britton, Hua~Rong Lu, Jutta Rohrbacher, An~N. Hermans,
  David~J. Gallacher, Robert~J.H. Greig, Alfonso {Bueno-Orovio}, and Blanca
  Rodriguez.
\newblock Human in silico drug trials demonstrate higher accuracy than animal
  models in predicting clinical pro-arrhythmic cardiotoxicity.
\newblock \emph{Frontiers in Physiology}, 8\penalty0 (SEP), 2017.
\newblock ISSN 1664042X.
\newblock \doi{10.3389/fphys.2017.00668}.

\bibitem[Pennington et~al.(1970)Pennington, Taylor, and
  Lown]{PenTayLow:1970:ctr}
James~E. Pennington, Jack Taylor, and Bernard Lown.
\newblock Chest {{Thump}} for {{Reverting Ventricular Tachycardia}}.
\newblock \emph{New England Journal of Medicine}, 283\penalty0 (22):\penalty0
  1192--1195, November 1970.
\newblock ISSN 0028-4793, 1533-4406.
\newblock \doi{10.1056/NEJM197011262832204}.

\bibitem[Perkins et~al.(2021)Perkins, Gr{\"a}sner, Semeraro, Olasveengen, Soar,
  Lott, Van De~Voorde, Madar, Zideman, Mentzelopoulos, Bossaert, Greif,
  Monsieurs, Svavarsd{\'o}ttir, Nolan, Ainsworth, Akin, Alfonzo, Andres,
  Attard~Montalto, Barelli, Baubin, Behringer, Bein, Biarent, Bingham, Blom,
  Boccuzzi, Borra, Bossaert, B{\"o}ttiger, Breckwoldt, Brissaud, Burkart,
  Cariou, Carli, Carmona, Cassan, Castren, Christophides, Cimpoesu, Clarens,
  Conaghan, Couper, Cronberg, De~Buck, De~Lucas, De~Roovere, Deakin, Delchef,
  Dirks, Djakow, Djarv, Druwe, Eldin, Ersdal, Friberg, Genbrugge, Georgiou,
  Goemans, {Gonzalez-Salvado}, Gradisek, Graesner, Greif, Handley, Hassager,
  Haywood, Heltne, Hendrickx, Herlitz, Hinkelbein, Hoffmann, Hunyadi~Anticevic,
  Johannesdottir, Khalifa, Klaassen, Koppl, Kreimeier, Kuzovlev, Lauritsen,
  Lilja, Lippert, Lockey, Lott, Lulic, Maas, Maconochie, Madar,
  {Martinez-Mejias}, Masterson, Mentzelopoulos, Meyran, Monsieurs, Morley,
  Moulaert, Mpotos, Nikolaou, Nolan, Olasveengen, Oliver, Paal, Pellis,
  Perkins, {Pflanzl-Knizacek}, Pitches, Poole, Raffay, Renier, Ristagno, Roehr,
  {Rosell-Ortiz}, Rudiger, Safri, Sanchez~Santos, Sandroni, Sari, Scapigliati,
  Schilder, Schlieber, Schnaubelt, Semeraro, Shammet, Singletary, Skare,
  Skrifvars, Smyth, Soar, Svavarsdottir, Szczapa, Taccone, Tageldin~Mustafa,
  Te~Pas, Thies, Tjelmeland, Trevisanuto, Truhlar, Trummer, Turner,
  Urlesberger, Vaahersalo, Van De~Voorde, Van~Grootven, Wilkinson, Wnent,
  Wyllie, Yeung, and
  Zideman]{PerGraSemOlaSoaLotVanMadZidMenBosGreMonSvaNolAinAkiAlfAndAttBarBauBehBeiBiaBinBloBocBorBosBotBreBriBurCarCarCarCasCasChrCimClaConCouCroDeDeDeDeaDelDirDjaDjaDruEldErsFriGenGeoGoeGonGraGraGreHanHasHayHelHenHerHinHofHunJohKhaKlaKopKreKuzLauLilLipLocLotLulMaaMacMadMarMasMenMeyMonMorMouMpoNikNolOlaOliPaaPelPerPflPitPooRafRenRisRoeRosRudSafSanSanSarScaSchSchSchSemShaSinSkaSkrSmySoaSvaSzcTacTagTeThiTjeTreTruTruTurUrlVaaVanVanWilWneWylYeuZid:2021:erc}
Gavin~D. Perkins, Jan-Thorsen Gr{\"a}sner, Federico Semeraro, Theresa
  Olasveengen, Jasmeet Soar, Carsten Lott, Patrick Van De~Voorde, John Madar,
  David Zideman, Spyridon Mentzelopoulos, Leo Bossaert, Robert Greif, Koen
  Monsieurs, Hildigunnur Svavarsd{\'o}ttir, Jerry~P. Nolan, S.~Ainsworth,
  S.~Akin, A.~Alfonzo, J.~Andres, S.~Attard~Montalto, A.~Barelli, M.~Baubin,
  W.~Behringer, B.~Bein, D.~Biarent, R.~Bingham, M.~Blom, A.~Boccuzzi,
  V.~Borra, L.~Bossaert, B.W. B{\"o}ttiger, J.~Breckwoldt, O.~Brissaud,
  R.~Burkart, A.~Cariou, P.~Carli, F.~Carmona, P.~Cassan, M.~Castren,
  T.~Christophides, C.D. Cimpoesu, C.~Clarens, P.~Conaghan, K.~Couper,
  T.~Cronberg, E.~De~Buck, N.~De~Lucas, A.~De~Roovere, C.D. Deakin, J.~Delchef,
  B.~Dirks, J.~Djakow, T.~Djarv, P.~Druwe, G.~Eldin, H.~Ersdal, H.~Friberg,
  C.~Genbrugge, M.~Georgiou, E.~Goemans, V.~{Gonzalez-Salvado}, P.~Gradisek,
  J.T. Graesner, R.~Greif, A.J. Handley, C.~Hassager, K.~Haywood, J.K. Heltne,
  D.~Hendrickx, J.~Herlitz, J.~Hinkelbein, F.~Hoffmann, S.~Hunyadi~Anticevic,
  G.B. Johannesdottir, G.~Khalifa, B.~Klaassen, J.~Koppl, U.~Kreimeier,
  A.~Kuzovlev, T.~Lauritsen, G.~Lilja, F.~Lippert, A.~Lockey, C.~Lott,
  I.~Lulic, M.~Maas, I.~Maconochie, J.~Madar, A.~{Martinez-Mejias},
  S.~Masterson, S.D. Mentzelopoulos, D.~Meyran, K.G. Monsieurs, C.~Morley,
  V.R.M. Moulaert, N.~Mpotos, N.~Nikolaou, J.P. Nolan, T.M. Olasveengen,
  E.~Oliver, P.~Paal, T.~Pellis, G.D. Perkins, L.~{Pflanzl-Knizacek},
  K.~Pitches, K.~Poole, V.~Raffay, W.~Renier, G.~Ristagno, C.C. Roehr,
  F.~{Rosell-Ortiz}, M.~Rudiger, A.~Safri, L.~Sanchez~Santos, C.~Sandroni,
  F.~Sari, A.~Scapigliati, S.~Schilder, J.~Schlieber, S.~Schnaubelt,
  F.~Semeraro, S.~Shammet, E.M. Singletary, C.~Skare, M.B. Skrifvars, M.~Smyth,
  J.~Soar, H.~Svavarsdottir, T.~Szczapa, F.~Taccone, M.~Tageldin~Mustafa,
  A.~Te~Pas, K.C. Thies, I.B.M. Tjelmeland, D.~Trevisanuto, A.~Truhlar,
  G.~Trummer, N.M. Turner, B.~Urlesberger, J.~Vaahersalo, P.~Van De~Voorde,
  H.~Van~Grootven, D.~Wilkinson, J.~Wnent, J.P. Wyllie, J.~Yeung, and D.A.
  Zideman.
\newblock European {{Resuscitation Council Guidelines}} 2021: {{Executive}}
  summary.
\newblock \emph{Resuscitation}, 161:\penalty0 1--60, April 2021.
\newblock ISSN 03009572.
\newblock \doi{10.1016/j.resuscitation.2021.02.003}.

\bibitem[Perotti et~al.(2015)Perotti, Krishnamoorthi, Borgstrom, Ennis, and
  Klug]{PerKriBorEnnKlu:2015:rsv}
L~E Perotti, S~Krishnamoorthi, N~P Borgstrom, D~B Ennis, and W~S Klug.
\newblock Regional segmentation of ventricular models to achieve repolarization
  dispersion in cardiac electrophysiology modeling.
\newblock \emph{International journal for numerical methods in biomedical
  engineering}, 31\penalty0 (8), August 2015.
\newblock ISSN 2040-7947.
\newblock \doi{10.1002/cnm.2718}.

\bibitem[Peyronnet et~al.(2016)Peyronnet, Nerbonne, and
  Kohl]{PeyNerKoh:2016:cmi}
R{\'e}mi Peyronnet, Jeanne~M. Nerbonne, and Peter Kohl.
\newblock Cardiac {{Mechano-Gated Ion Channels}} and {{Arrhythmias}}.
\newblock \emph{Circulation Research}, 118\penalty0 (2):\penalty0 311--329,
  January 2016.
\newblock ISSN 0009-7330, 1524-4571.
\newblock \doi{10.1161/CIRCRESAHA.115.305043}.

\bibitem[Ponnaluri et~al.(2016)Ponnaluri, Perotti, Liu, Qu, Weiss, Ennis, Klug,
  and Garfinkel]{PonPerLiuQuWeiEnnKluGar:2016:ehf}
Aditya V.~S. Ponnaluri, Luigi~E. Perotti, Michael Liu, Zhilin Qu, James~N.
  Weiss, Daniel~B. Ennis, William~S. Klug, and Alan Garfinkel.
\newblock Electrophysiology of {{Heart Failure Using}} a {{Rabbit Model}}:
  {{From}} the {{Failing Myocyte}} to {{Ventricular Fibrillation}}.
\newblock \emph{PLOS Computational Biology}, 12\penalty0 (6):\penalty0
  e1004968, June 2016.
\newblock ISSN 1553-7358.
\newblock \doi{10.1371/journal.pcbi.1004968}.

\bibitem[Quinn and Kohl(2016)]{QuiKoh:2016:cmr}
T.~Alexander Quinn and Peter Kohl.
\newblock Comparing maximum rate and sustainability of pacing by mechanical vs.
  electrical stimulation in the {{Langendorff-perfused}} rabbit heart.
\newblock \emph{EP Europace}, 18\penalty0 (suppl\_4):\penalty0 iv85--iv93,
  December 2016.
\newblock ISSN 1099-5129, 1532-2092.
\newblock \doi{10.1093/europace/euw354}.

\bibitem[Quinn et~al.(2017)Quinn, Jin, Lee, and Kohl]{QuiJinLeeKoh:2017:mie}
T.~Alexander Quinn, Honghua Jin, Peter Lee, and Peter Kohl.
\newblock Mechanically {{Induced Ectopy}} via {{Stretch-Activated
  Cation-Nonselective Channels Is Caused}} by {{Local Tissue Deformation}} and
  {{Results}} in {{Ventricular Fibrillation}} if {{Triggered}} on the
  {{Repolarization Wave Edge}} ({{Commotio Cordis}}).
\newblock \emph{Circulation: Arrhythmia and Electrophysiology}, 10\penalty0
  (8):\penalty0 e004777, August 2017.
\newblock \doi{10.1161/CIRCEP.116.004777}.

\bibitem[Rackauckas and Nie(2017)]{RacNie:2017:djp}
Christopher Rackauckas and Qing Nie.
\newblock {{DifferentialEquations}}.jl -- {{A Performant}} and {{Feature-Rich
  Ecosystem}} for {{Solving Differential Equations}} in {{Julia}}.
\newblock \emph{Journal of Open Research Software}, 5\penalty0 (1):\penalty0
  15, May 2017.
\newblock ISSN 2049-9647.
\newblock \doi{10.5334/jors.151}.

\bibitem[Ranade et~al.(2014)Ranade, Qiu, Woo, Hur, Murthy, Cahalan, Xu, Mathur,
  Bandell, Coste, Li, Chien, and
  Patapoutian]{RanQiuWooHurMurCahXuMatBanCosLiChiPat:2014:pma}
Sanjeev~S. Ranade, Zhaozhu Qiu, Seung-Hyun Woo, Sung~Sik Hur, Swetha~E. Murthy,
  Stuart~M. Cahalan, Jie Xu, Jayanti Mathur, Michael Bandell, Bertrand Coste,
  Yi-Shuan~J. Li, Shu Chien, and Ardem Patapoutian.
\newblock Piezo1, a mechanically activated ion channel, is required for
  vascular development in mice.
\newblock \emph{Proceedings of the National Academy of Sciences}, 111\penalty0
  (28):\penalty0 10347--10352, July 2014.
\newblock ISSN 0027-8424, 1091-6490.
\newblock \doi{10.1073/pnas.1409233111}.

\bibitem[Reed et~al.(2014)Reed, Kohl, and Peyronnet]{ReeKohPey:2014:mcc}
Alistair Reed, Peter Kohl, and R{\'e}mi Peyronnet.
\newblock Molecular candidates for cardiac stretch-activated ion channels.
\newblock \emph{Global Cardiology Science \& Practice}, 2014\penalty0
  (2):\penalty0 9--25, June 2014.
\newblock ISSN 2305-7823.
\newblock \doi{10.5339/gcsp.2014.19}.

\bibitem[Sachs(2010)]{Sac:2010:sic}
Frederick Sachs.
\newblock Stretch-{{Activated Ion Channels}}: {{What Are They}}?
\newblock \emph{Physiology}, 25\penalty0 (1):\penalty0 50--56, February 2010.
\newblock ISSN 1548-9213, 1548-9221.
\newblock \doi{10.1152/physiol.00042.2009}.

\bibitem[S{\'a}nchez et~al.(2018)S{\'a}nchez, D'Ambrosio, Maffessanti, Caiani,
  Prinzen, Krause, Auricchio, and Potse]{SanDAMafCaiPriKraAurPot:2018:sav}
C.~S{\'a}nchez, G.~D'Ambrosio, F.~Maffessanti, E.~G. Caiani, F.~W. Prinzen,
  R.~Krause, A.~Auricchio, and M.~Potse.
\newblock Sensitivity analysis of ventricular activation and electrocardiogram
  in tailored models of heart-failure patients.
\newblock \emph{Medical \& Biological Engineering \& Computing}, 56\penalty0
  (3):\penalty0 491--504, March 2018.
\newblock \doi{10.1007/s11517-017-1696-9}.

\bibitem[Saotome et~al.(2018)Saotome, Murthy, Kefauver, Whitwam, Patapoutian,
  and Ward]{SaoMurKefWhiPatWar:2018:sma}
Kei Saotome, Swetha~E. Murthy, Jennifer~M. Kefauver, Tess Whitwam, Ardem
  Patapoutian, and Andrew~B. Ward.
\newblock Structure of the mechanically activated ion channel {{Piezo1}}.
\newblock \emph{Nature}, 554\penalty0 (7693):\penalty0 481--486, February 2018.
\newblock ISSN 0028-0836, 1476-4687.
\newblock \doi{10.1038/nature25453}.

\bibitem[Schott(1920)]{Sch:1920:vss}
Eduard~Franz Schott.
\newblock On ventricular standstill [{{Stokes}}-{{Adams}} attacks] and other
  arrhythmias of temporary nature.
\newblock page 211, 1920.

\bibitem[Simon-Chica et~al.(2024)Simon-Chica, Klesen, Emig, Chan, Greiner,
  Gr{\"u}n, Lother, Hilgendorf, Rog-Zielinska, Ravens, Kohl, Schneider-Warme,
  and Peyronnet]{SimKleEmiChaGreGruLotHilRogRavKohSchPey:2024:psc}
Ana Simon-Chica, Alexander Klesen, Ramona Emig, Andy Chan, Joachim Greiner,
  Dominic Gr{\"u}n, Achim Lother, Ingo Hilgendorf, Eva~A. Rog-Zielinska, Ursula
  Ravens, Peter Kohl, Franziska Schneider-Warme, and R{\'e}mi Peyronnet.
\newblock Piezo1 stretch-activated channel activity differs between murine bone
  marrow-derived and cardiac tissue-resident macrophages.
\newblock \emph{The Journal of Physiology}, 602\penalty0 (18):\penalty0
  4437--4456, September 2024.
\newblock ISSN 0022-3751, 1469-7793.
\newblock \doi{10.1113/JP284805}.

\bibitem[Su et~al.(2023)Su, Zhang, Li, Xi, Lu, Shen, Ma, Wang, Shen, Xie, Ma,
  Xie, and Xiang]{SuZhaLiXiLuSheMaWanSheXieMaXieXia:2023:cpe}
Sheng-an Su, Yuhao Zhang, Wudi Li, Yutao Xi, Yunrui Lu, Jian Shen, Yuankun Ma,
  Yaping Wang, Yimin Shen, Lan Xie, Hong Ma, Yao Xie, and Meixiang Xiang.
\newblock Cardiac {{Piezo1 Exacerbates Lethal Ventricular Arrhythmogenesis}} by
  {{Linking Mechanical Stress}} with {{Ca}}{\textsuperscript{2+}} {{Handling
  After Myocardial Infarction}}.
\newblock \emph{Research}, 6:\penalty0 0165, January 2023.
\newblock ISSN 2639-5274.
\newblock \doi{10.34133/research.0165}.

\bibitem[Syeda et~al.(2015)Syeda, Xu, Dubin, Coste, Mathur, Huynh, Matzen, Lao,
  Tully, Engels, Petrassi, Schumacher, Montal, Bandell, and
  Patapoutian]{SyeXuDubCosMatHuyMatLaoTulEngPetSchMonBanPat:2015:cam}
Ruhma Syeda, Jie Xu, Adrienne~E Dubin, Bertrand Coste, Jayanti Mathur, Truc
  Huynh, Jason Matzen, Jianmin Lao, David~C Tully, Ingo~H Engels, H~Michael
  Petrassi, Andrew~M Schumacher, Mauricio Montal, Michael Bandell, and Ardem
  Patapoutian.
\newblock Chemical activation of the mechanotransduction channel {{Piezo1}}.
\newblock \emph{eLife}, 4:\penalty0 e07369, May 2015.
\newblock ISSN 2050-084X.
\newblock \doi{10.7554/eLife.07369}.

\bibitem[Tomek et~al.(2019)Tomek, {Bueno-Orovio}, Passini, Zhou, Minchole,
  Britton, Bartolucci, Severi, Shrier, Virag, Varro, and
  Rodriguez]{TomBuePasZhoMinBriBarSevShrVirVarRod:2019:dcv}
Jakub Tomek, Alfonso {Bueno-Orovio}, Elisa Passini, Xin Zhou, Ana Minchole,
  Oliver Britton, Chiara Bartolucci, Stefano Severi, Alvin Shrier, Laszlo
  Virag, Andras Varro, and Blanca Rodriguez.
\newblock Development, calibration, and validation of a novel human ventricular
  myocyte model in health, disease, and drug block.
\newblock \emph{eLife}, 8:\penalty0 e48890, December 2019.
\newblock ISSN 2050-084X.
\newblock \doi{10.7554/eLife.48890}.

\bibitem[Wall et~al.(2012)Wall, Guccione, Ratcliffe, and
  Sundnes]{WalGucRatSun:2012:efr}
Samuel~T. Wall, Julius~M. Guccione, Mark~B. Ratcliffe, and Joakim~S. Sundnes.
\newblock Electromechanical feedback with reduced cellular connectivity alters
  electrical activity in an infarct injured left ventricle: A finite element
  model study.
\newblock \emph{American Journal of Physiology-Heart and Circulatory
  Physiology}, 302\penalty0 (1):\penalty0 H206--H214, January 2012.
\newblock ISSN 0363-6135, 1522-1539.
\newblock \doi{10.1152/ajpheart.00272.2011}.

\bibitem[Wijerathne et~al.(2022)Wijerathne, Ozkan, and
  Lacroix]{WijOzkLac:2022:yef}
Tharaka~D. Wijerathne, Alper~D. Ozkan, and J{\'e}r{\^o}me~J. Lacroix.
\newblock Yoda1's energetic footprint on {{Piezo1}} channels and its modulation
  by voltage and temperature.
\newblock \emph{Proceedings of the National Academy of Sciences}, 119\penalty0
  (29):\penalty0 e2202269119, July 2022.
\newblock ISSN 0027-8424, 1091-6490.
\newblock \doi{10.1073/pnas.2202269119}.

\bibitem[Wu et~al.(2017)Wu, Young, Lewis, Martfeld, Kalmeta, and
  Grandl]{WuYouLewMarKalGra:2017:ima}
Jason Wu, Michael Young, Amanda~H. Lewis, Ashley~N. Martfeld, Breanna Kalmeta,
  and J{\"o}rg Grandl.
\newblock Inactivation of {{Mechanically Activated Piezo1 Ion Channels Is
  Determined}} by the {{C-Terminal Extracellular Domain}} and the {{Inner Pore
  Helix}}.
\newblock \emph{Cell Reports}, 21\penalty0 (9):\penalty0 2357--2366, November
  2017.
\newblock ISSN 22111247.
\newblock \doi{10.1016/j.celrep.2017.10.120}.

\bibitem[Yu et~al.(2022{\natexlab{a}})Yu, Gong, Kesteven, Guo, Wu, Li, Iismaa,
  Kaidonis, Graham, Cox, Feneley, and
  Martinac]{YuGonKesGuoWuLiIisKaiGraCoxFenMar:2022:ptw}
Ze-Yan Yu, Hutao Gong, Scott Kesteven, Yang Guo, Jianxin Wu, Jinyuan Li, Siiri
  Iismaa, Xenia Kaidonis, Robert~M. Graham, Charles~D. Cox, Michael~P. Feneley,
  and Boris Martinac.
\newblock Piezo1 and {{TRPM4}} work in tandem to initiate cardiac hypertrophic
  signalling in response to pressure overload.
\newblock \emph{Biophysical Journal}, 121\penalty0 (3):\penalty0 493a, February
  2022{\natexlab{a}}.
\newblock ISSN 00063495.
\newblock \doi{10.1016/j.bpj.2021.11.321}.

\bibitem[Yu et~al.(2022{\natexlab{b}})Yu, Gong, Kesteven, Guo, Wu, Li, Cheng,
  Zhou, Iismaa, Kaidonis, Graham, Cox, Feneley, and
  Martinac]{YuGonKesGuoWuLiCheZhoIisKaiGraCoxFenMar:2022:pcm}
Ze-Yan Yu, Hutao Gong, Scott Kesteven, Yang Guo, Jianxin Wu, Jinyuan~Vero Li,
  Delfine Cheng, Zijing Zhou, Siiri~E. Iismaa, Xenia Kaidonis, Robert~M.
  Graham, Charles~D. Cox, Michael~P. Feneley, and Boris Martinac.
\newblock Piezo1 is the cardiac mechanosensor that initiates the cardiomyocyte
  hypertrophic response to pressure overload in adult mice.
\newblock \emph{Nature Cardiovascular Research}, 1\penalty0 (6):\penalty0
  577--591, June 2022{\natexlab{b}}.
\newblock ISSN 2731-0590.
\newblock \doi{10.1038/s44161-022-00082-0}.

\bibitem[Yue et~al.(2015)Yue, Xie, Yu, Stock, Du, and
  Yue]{YueXieYuStoDuYue:2015:rtc}
Zhichao Yue, Jia Xie, Albert~S. Yu, Jonathan Stock, Jianyang Du, and Lixia Yue.
\newblock Role of {{TRP}} channels in the cardiovascular system.
\newblock \emph{American Journal of Physiology-Heart and Circulatory
  Physiology}, 308\penalty0 (3):\penalty0 H157--H182, February 2015.
\newblock ISSN 0363-6135, 1522-1539.
\newblock \doi{10.1152/ajpheart.00457.2014}.

\bibitem[Zarychanski et~al.(2012)Zarychanski, Schulz, Houston, Maksimova,
  Houston, Smith, Rinehart, and Gallagher]{ZarSchHouMakHouSmiRinGal:2012:mmp}
Ryan Zarychanski, Vincent~P. Schulz, Brett~L. Houston, Yelena Maksimova,
  Donald~S. Houston, Brian Smith, Jesse Rinehart, and Patrick~G. Gallagher.
\newblock Mutations in the mechanotransduction protein {{PIEZO1}} are
  associated with hereditary xerocytosis.
\newblock \emph{Blood}, 120\penalty0 (9):\penalty0 1908--1915, August 2012.
\newblock ISSN 0006-4971, 1528-0020.
\newblock \doi{10.1182/blood-2012-04-422253}.

\bibitem[Zhang et~al.(2017)Zhang, Chi, Jiang, Zhao, and
  Xiao]{ZhaChiJiaZhaXia:2017:pim}
Tingxin Zhang, Shaopeng Chi, Fan Jiang, Qiancheng Zhao, and Bailong Xiao.
\newblock A protein interaction mechanism for suppressing the mechanosensitive
  {{Piezo}} channels.
\newblock \emph{Nature Communications}, 8\penalty0 (1):\penalty0 1797, December
  2017.
\newblock ISSN 2041-1723.
\newblock \doi{10.1038/s41467-017-01712-z}.

\bibitem[Zhang and Zou(2022)]{ZhaZou:2022:ntm}
Yu~Zhang and Qingsong Zou.
\newblock A {{Novel Triple-gate Model}} for {{Mechanosensitive Ion Channel
  Piezo1}}.
\newblock In \emph{2022 10th {{International Conference}} on {{Bioinformatics}}
  and {{Computational Biology}} ({{ICBCB}})}, pages 135--141, Hangzhou, China,
  May 2022. IEEE.
\newblock ISBN 978-1-66540-108-1.
\newblock \doi{10.1109/ICBCB55259.2022.9802499}.

\bibitem[Zhang et~al.(2021)Zhang, Su, Li, Ma, Shen, Wang, Shen, Chen, Ji, Xie,
  Ma, and Xiang]{ZhaSuLiMaSheWanSheCheJiXieMaXia:2021:pmp}
Yuhao Zhang, Sheng-an Su, Wudi Li, Yuankun Ma, Jian Shen, Yaping Wang, Yimin
  Shen, Jian Chen, Yongli Ji, Yao Xie, Hong Ma, and Meixiang Xiang.
\newblock Piezo1-{{Mediated Mechanotransduction Promotes Cardiac Hypertrophy}}
  by {{Impairing Calcium Homeostasis}} to {{Activate Calpain}}/{{Calcineurin
  Signaling}}.
\newblock \emph{Hypertension}, 78\penalty0 (3):\penalty0 647--660, September
  2021.
\newblock ISSN 0194-911X, 1524-4563.
\newblock \doi{10.1161/HYPERTENSIONAHA.121.17177}.

\bibitem[Zheng et~al.(2019)Zheng, Gracheva, and
  Bagriantsev]{ZheGraBag:2019:hgi}
Wang Zheng, Elena~O Gracheva, and Sviatoslav~N Bagriantsev.
\newblock A hydrophobic gate in the inner pore helix is the major determinant
  of inactivation in mechanosensitive {{Piezo}} channels.
\newblock \emph{eLife}, 8:\penalty0 e44003, January 2019.
\newblock ISSN 2050-084X.
\newblock \doi{10.7554/eLife.44003}.

\bibitem[Zheng and Sachs(2017)]{ZheSac:2017:isd}
Wenjun Zheng and Frederick Sachs.
\newblock Investigating the structural dynamics of the {{PIEZO1}} channel
  activation and inactivation by coarse-grained modeling.
\newblock \emph{Proteins: Structure, Function, and Bioinformatics}, 85\penalty0
  (12):\penalty0 2198--2208, December 2017.
\newblock ISSN 0887-3585, 1097-0134.
\newblock \doi{10.1002/prot.25384}.

\end{thebibliography}

\end{document}